\documentclass[12pt]{amsart}
\theoremstyle{plain}
\numberwithin{equation}{section}
\newtheorem{Theorem}{Theorem}
\newtheorem{Lemma}[Theorem]{Lemma}

\newtheorem{Proposition}[Theorem]{Proposition}

\theoremstyle{remark}

\title[Hill operators with a two term potential]
{Asymptotics of instability zones of the Hill operator with
a two term potential}

\author{Plamen Djakov}

\author{Boris Mityagin}

\begin{document}
\pagestyle{myheadings}
\markboth{}{}

\address{Department of Mathematics,
Sofia University,
1164 Sofia, Bulgaria}

 \email{djakov@fmi.uni-sofia.bg}

\address{Department of Mathematics,
The Ohio State University,
 231 West 18th Ave,
Columbus, OH 43210, USA}
\email{mityagin.1@osu.edu}

\begin{abstract}
Let $\gamma_n $ denote the length of the $n$-th zone of instability
of the Hill operator
$Ly= -y^{\prime \prime} - [4t\alpha \cos2x + 2 \alpha^2 \cos 4x ] y,
 $
where $\alpha \neq 0, $ and either both $\alpha, t $ are
real, or both are pure imaginary numbers.
For even $n$ we prove: if $t, n $ are fixed, then for
$ \alpha \to 0 $
$$
\gamma_n =\left | \frac{8\alpha^n}{2^n [(n-1)!]^2} \prod_{k=1}^{n/2}
\left ( t^2 - (2k-1)^2 \right ) \right | \left ( 1 + O(\alpha) \right ),
$$
and if $ \alpha, t $  are fixed, then for
$ n \to \infty $  
$$
\gamma_n =  \frac{8 |\alpha/2|^n}{[2 \cdot 4 \cdots (n-2)]^2}
\left | \cos \left (\frac{\pi}{2} t \right ) \right |
\left [ 1 + O \left (\frac{\log n}{n} \right ) \right ].
$$
Similar formulae (see Theorems \ref{thm2} and \ref{thm4}) hold for odd $n.$
The asymptotics for $\alpha \to 0 $ imply
interesting identities for squares of integers
(see Sect. 4, Thm.\ref{thm3}).

\end{abstract}

\maketitle

\section{Introduction. Main Results.}

The Schr\"odinger operator, considered on $ \mathbb{R},$
\begin{equation}
\label{i01}
Ly = - y^{\prime \prime} + v(x) y ,
\end{equation}
with a real-valued periodic $L^2 
([0,\pi])$-potential $v(x), $
$ v(x+\pi) = v(x), $
has spectral gaps, or instability zones, $(\lambda^-_n, \lambda^+_n),$
$ n \geq 1 ,$ close to $n^2 $ if $n$ is large enough.
The points $\lambda^-_n, \lambda^+_n $
could be determined as eigenvalues of the Hill equation
\begin{equation}
\label{i02}
 - y^{\prime \prime} + v(x) y   
= \lambda y,
\end{equation}
considered on $[0,\pi]$ with boundary conditions
\begin{equation}
\label{i03}
Per^+: \quad y(0) = y(\pi), \quad y^\prime (0) = y^\prime (\pi),
\end{equation}
for even $n,$ and
\begin{equation}
\label{i04}
Per^+: \quad y(0) =- y(\pi), \quad y^\prime (0) = - y^\prime (\pi), 
\end{equation}
for odd $n.$
See basics and details in \cite{E,LS,MW,Mar,PT}.

The rate of decay of the sequence of spectral gaps
$\gamma_n = \lambda_n^+ - \lambda_n^- $
is closely related to the smoothness of the
potential $v.$
We'll mention now only
Hochshtadt's result \cite{Hoch63} that
an $L^2 ([0,\pi])$-potential $v$ is in $C^\infty $ if and only if
$(\gamma_n )$ decays faster than any power of $1/n, $
and Trubowitz's result \cite{Tr} that
an $L^2 ([0,\pi])$-potential $v$ is analytic if and only if
$(\gamma_n )$ decays exponentially.
See further references and later results in
\cite{DM3,DM5}.

In the case of specific potentials,
like the Mathieu potential
\begin{equation}
\label{i05}
v(x) = 2a \cos 2x, \quad a\neq 0, \;\text{real},
\end{equation}
or more general trigonometric polynomials
\begin{equation}
\label{i06}
v(x) = \sum_{-N}^N c_k \exp (ikx), \quad c_k = \overline{c_{-k}},
\quad 0 \leq k \leq N < \infty,
\end{equation}
one comes to two classes of questions:

(i) Is the $n$-th zone of instability closed, i.e.,
\begin{equation}
\label{i07}
\gamma_n = \lambda_n^+ - \lambda_n^- = 0,
\end{equation}
or, equivalently, is the multiplicity of $\lambda_n^+ $
equal to 2?

(ii) If $ \gamma_n \neq 0, $ could
we tell more about the size of this gap,
or, for large enough $n,$ what is the asymptotic behavior of
$\gamma_n = \gamma_n (v)?$

E. L. Ince \cite{I} answered in a negative way 
Question (i) in the case of the potential (\ref{i05}):
the Mathieu-Hill operator has only {\em simple }
eigenvalues both for $Per^+ $ and $ Per^- $ boundary conditions,
i.e., all zones of instability of the Mathieu--Schr\"odinger operator
are open. His proof is presented in \cite{E};
see other proofs of this fact
in \cite{Hil,Mark,McL}, and further references in \cite{E,MW58}.

For fixed $n$ and $a \to 0, $ D. Levy and J. Keller \cite{LK}
gave the asymptotics of $ \gamma_n = \gamma_n (a),\; a \in (\ref{i05});$
namely
\begin{equation}
\label{i08}
\gamma_n = \lambda_n^+ - \lambda_n^- =
\frac{8(|a|/4)^n}{[(n-1)!]^2} \left ( 1+ O(a) \right ),
\end{equation}

Almost 20 years later, E. Harrell \cite{Har} found,
up to a constant factor, 
the asymptotics of the spectral gaps of
the Mathieu potential (\ref{i05})
as $ n\to \infty. $
J. Avron and B. Simon \cite{AS} gave an alternative proof of
E. Harrell's asymptotics 
and found the exact value of the constant factor,
which led to the following formula
\begin{equation}
\label{i09}
\gamma_n = \lambda_n^+ - \lambda_n^- =
\frac{8(|a|/4)^n}{[(n-1)!]^2} \left ( 1+
O\left (\frac{1}{n^2}\right ) \right ).
\end{equation}
Let us mention that in \cite{DM8,DM9} we found the asymptotics of
the spectral gaps of 1D Dirac operator with cosine potential
and multiplicities of {\em all} periodic and antiperiodic eigenvalues.

J. Avron and B. Simon \cite{AS}
raised the question about the asymptotics of spectral gaps 
in the case of a two term potential
\begin{equation}
\label{i10}
v(x) = a \cos 2x + b \cos 4x.
\end{equation}
A. Grigis \cite{Gr} asked essentially the same question
for the isospectral potential
$ v(x) = a \sin 2x - b \cos 4x.$

We found such asymptotics.
Our results (see below) are announced in \cite{DM11}, and
the present paper gives their detailed proofs.

Put 
for real $a, b\neq 0 $
\begin{equation}
\label{i11}
a = - 4 \alpha t, \quad b = - 2 \alpha^2,
\end{equation}
where either

(\ref{i11}a)  both $\alpha $ and $t$ are real (if $b<0$)

or

(\ref{i11}b) both $\alpha $ and $t$ are pure imaginary  (if $b>0 $).

This special parametrization comes from the Magnus-Winkler analysis
\cite{MW,MW58} of this Hill operator [{\em Whittaker operator} in their 
terminology].
Our paper \cite{DM12} is essentially an algebraic introduction
to the present article. In \cite{DM12}
we sharpen the Magnus-Winkler result on existence of finitely
many zones of instability
in the case of the integers
$t$ in (\ref{i11}).
The special role of integer
$t$'s is incorporated into coefficients
in the asymptotics of $\gamma_n (\alpha), \; \alpha \to 0, $
and $ \gamma_n (v), \; n\to \infty, $ with $v \in (\ref{i10})+(\ref{i11}).$
Namely (see Thm. \ref{thm2}),
if $t$ and $n$ are fixed, then
for even n
\begin{equation}
\label{i13}
\gamma_n = \left | \frac{8\alpha^n}{2^n [(n-1)!]^2} \prod_{k=1}^{n/2}
\left ( t^2 - (2k-1)^2 \right ) \right |
\left ( 1 + O(\alpha) \right ),
\end{equation}
and for odd $n$
\begin{equation}
\label{i14}
\gamma_n = \left | \frac{ 8\alpha^n t}{2^n [(n-1)!]^2}
\prod_{k=1}^{(n-1)/2}
\left ( t^2 - (2k)^2 \right ) \right | \left ( 1 + O(\alpha ) \right ).
\end{equation}

This could be compared with
the Levy-Keller statement (\ref{i08}) above.

If  $ \alpha \neq 0, \, t \neq 0 $ are fixed,
then (see Thm. \ref{thm4})
the following asymptotic formulae hold 
as $n \to \infty:$
for even $n$
\begin{equation}
\label{i15}
\gamma_n =  \frac{8 |\alpha/2|^n}{[2 \cdot 4 \cdots (n-2)]^2}
\left | \cos \left (\frac{\pi}{2} t \right ) \right |
\left [ 1 + O \left (\frac{\log n}{n} \right ) \right ],
\end{equation}
and for odd $n$
\begin{equation}
\label{i16}
\gamma_n =
 \frac{8 |\alpha/2|^n}{[1 \cdot 3 \cdots (n-2)]^2}
\frac{2}{\pi} \left | \sin \left ( \frac{\pi}{2} t \right )  \right |
\left [ 1 + O \left (\frac{\log n}{n} \right ) \right ].
\end{equation}

This result could be compared with
the Harrell--Avron--Simon formula
(\ref{i09}) above.

Asymptotics (\ref{i13})-(\ref{i14}) imply
interesting identities for squares of integers
(see Sect. 4, Thm.\ref{thm3}).

Our proofs are based on an almost explicit formula
(see Thm. \ref{thm1}) 
for $\gamma_n $
in terms of Fourier coefficients of the potential
$$
v(x) = \sum_{k\; even} V_k e^{ikx}, \quad x \in [0,2\pi].
$$

We proved it in \cite{DM3}, Thm. 8.
For convenience, we give in Sect. 2 all
details to adjust this formula to both cases
$\alpha \to 0 \, $    $(n,t $ fixed), and
$n \to \infty $
($\alpha, t$ fixed).

\section{Preliminaries}

1. Let $v $
be a periodic function of a period $\pi.$
The differential operator
\begin{equation}
\label{b1}
Ly = - y^{\prime \prime} + v(x)y, \quad x \in [0,\pi],
\end{equation}
considered with periodic boundary conditions
$$ Per^+ :\quad  y(0) = y(\pi), \;\;
y^\prime (0) = y^\prime (\pi) ,$$
or antiperiodic boundary conditions
$$ Per^- :\quad  y(0) =- y(\pi), \;\;
y^\prime (0) = - y^\prime (\pi) ,$$
is known as the Hill operator with potential $v.$
It is self-adjoint for real-valued potentials.

Consider
the operator
$$L^0 y = - y^{\prime \prime}.
$$
The periodic and antiperiodic spectra of $L^0 $
are discrete, and we have
$$ \sigma_{Per^+} = \{n^2, \; n \; \mbox{even} \},
\qquad \sigma_{Per^-} = \{n^2, \; n\; \mbox{odd} \}.
$$
Moreover, each eigenvalue $n^2 \neq 0 $ is of multiplicity 2,
and
\begin{equation}
\label{b2}
e_{-n} = e^{-inx}, \quad  e_n = e^{inx}
\end{equation}
are eigenfunctions corresponding to $n^2.$
So, if we consider periodic boundary conditions, then
$ \lambda = 0 $ is the only eigenvalue of $L^0 $ of
multiplicity 1, and the constant function $e_0 = 1 $
is the  corresponding normalized eigenfunction.

If $L^2([0,\pi])$ is considered with the scalar product
$$ (f,g) = \frac{1}{\pi} \int_0^\pi f(x)\overline{g(x)} dx, $$
then each of the families of functions
$\{e_{2k}, k \in \mathbb{Z} \}$
and $\{e_{2k-1}, k \in \mathbb{Z} \}$
is an orthonormal basis in $L^2([0,\pi]) .$
The basis 
$\{e_{2k}, k \in \mathbb{Z} \}$
(respectively,
$\{e_{2k-1}, k \in \mathbb{Z} \}$)
is used when we study the periodic
(respectively, antiperiodic)
spectra of $L.$

We always assume that $ v \in L^2 ([0,\pi]) $ and
denote by $\|v\|$ its $L^2$-norm. Since $v$ has a period $\pi $
its
Fourier series can be written in the form
$$v(x) = \sum_{m \in \mathbb{Z}} V(m) \exp (imx), \qquad
V(m) = 0 \;\; \text{for odd} \; m;
$$
then $\|v\|^2 = \sum  |V(m)|^2.$

It is well known (see \cite{MW}, Thm 2.1, or \cite{E}, Thm 2.3.1)
that the periodic and antiperiodic spectra of $L$ are discrete,
and moreover, there is a sequence of real numbers
\begin{equation}
\label{b3}
\lambda_0 < \lambda^-_1  \leq \lambda^+_1 <
\lambda^-_2  \leq \lambda^+_2 <
\lambda^-_3  \leq \lambda^+_3 <
\lambda^-_4  \leq \lambda^+_4 < \cdots
\end{equation}
such that the terms with even (respectively, odd) indices
give the periodic (respectively, antiperiodic) spectra of $L.$
We have
$$
\lambda_0 = 0, \quad   \lambda^-_n = \lambda^+_n = n^2, \;\;
\forall n \in \mathbb{N}
\quad \text{if}  \;\; v(x) \equiv 0.
$$
The Hill operator $L = L^0 + v $ may be regarded as a perturbation
of $L^0.$  A perturbation type argument shows that
if $\|v \| $ is small then $ \lambda_0 $ is close to 0,
and $\lambda^-_n,  \lambda^+_n $ are close to $n^2. $

The following proposition states and proves this fact in a more precise
form.
Of course, different versions of this statement
are well known but we give the version which is
convenient for our purposes.

\begin{Proposition}
\label{prop1}
(a) If $\|v\| \leq 1/4, $ then
\begin{equation}
\label{b4}
|\lambda_0 | \leq  4\|v\| \quad  \text{and} \quad
|\lambda_n^\pm - n^2 | \leq 4 \|v\| \quad \text{for} \;\; n \in \mathbb{N}.
\end{equation}

(b)  If $V(0) = \frac{1}{\pi} \int_0^\pi v(x) dx = 0, $
then there is a constant $N_0 = N_0 (v) $ such that
\begin{equation}
\label{b5}
|\lambda_n^\pm - n^2 | < 1 \quad \text{for} \;\; n \geq N_0.
\end{equation}

\end{Proposition}

\begin{proof}  A proof of Part (b) can be found in \cite{Mar},
Thm. 1.5.2, p. 76.
We need only the case when Thm. 1.5.2 claims, as it is observed there
on p. 241, (3.4.5), that if $V(0) = 0,$ i.e., $a_1 =0 $
in (3.4.5), then
\begin{equation}
\label{b6}
\lambda_n^\pm = n^2 + \varepsilon_n^\pm, \qquad
  E^2 = \sum_{n=0}^\infty |\varepsilon_n^\pm |^2 < \infty.
\end{equation}
If $N_0 $ is chosen so that
\begin{equation}
\label{b7}
\sum_{n= N_0}^\infty |\varepsilon_n^\pm |^2  <1,
\end{equation}
then of course (\ref{b6}) and (\ref{b7}) imply (\ref{b5}).
It is not said in \cite{Mar} explicitly
but, by following the proofs of Thm. 1.5.2 and its preliminaries,
one can get estimates of $E$ in terms of the norm
$\|v\| $ and explain (\ref{b4}), maybe
with other (absolute) constants, instead of 1/4 and 4.
To avoid any doubts (or careful reading of tens pages in \cite{Mar}),
we give an alternative proof of Proposition \ref{prop1} as an exercise
in Perturbation Theory.

We use Fourier analysis, by
considering the basis $(e_k)_{k \in 2\mathbb{Z}} $
in the periodic case, and the basis
$(e_k)_{k \in 2\mathbb{Z}-1} $
in the antiperiodic case.
Of course, each operator in $L^2 ([0,\pi]) $ is identified
with the corresponding operator in
$\ell^2 (2\mathbb{Z}) $ or
$\ell^2 (2\mathbb{Z}-1) $
(and with the corresponding matrix representation).

Let $V$ denote the operator
$y \to v(x)y, $
and let $ R^0_\lambda = (\lambda - L^0 )^{-1}.$
The matrix representations of $R^0_\lambda  $  and $V$ are                                       
$$
\left ( R^0_\lambda \right )_{km} =
\frac{1}{\lambda - m^2} \delta_{km}, \qquad
V_{km} = V(k-m).
$$
Since $L= L^0 + V $
we have
$$ \lambda - L = \lambda - L^0 - V = (\lambda - L^0)(1- R^0_\lambda V);
$$
thus,
whenever $\|R^0_\lambda V \| < 1, $
the operator
\begin{equation}
\label{b8}
R_\lambda = (\lambda - L)^{-1} = (1- R^0_\lambda V)^{-1} R^0_\lambda =
\sum_{k=0}^\infty (R^0_\lambda V)^k R^0_\lambda .
\end{equation}
is well defined.

The matrix representations of 
$R^0_\lambda V $ is
\begin{equation}
\label{b9}
\left (R^0_\lambda V \right )_{km} =
\frac{V(k-m)}{\lambda - k^2}. 
\end{equation}
Since the operator norm in $\ell^2 $-norm does not exceed the 
Hilbert-Schmidt norm,
we have
\begin{equation}
\label{b10}
\|R^0_\lambda V\|^2 \leq
\sum_{k} \frac{1}{|\lambda - k^2|^2} \cdot
\sum_{m} |V(k-m)|^2  = A(\lambda)
\cdot \|v\|^2.
\end{equation}
with $A(\lambda) = A^+ (\lambda) $ if $bc = Per^+, $
and $A(\lambda) = A^- (\lambda) $ if $bc = Per^-, $
where
$$
A^+ (\lambda) =
\sum_{k \in  2\mathbb{Z}} \frac{1}{|\lambda - k^2|^2},
\qquad
A^- (\lambda) =
\sum_{k \in  2\mathbb{Z} -1} \frac{1}{|\lambda - k^2|^2}.
$$

To estimate $A (\lambda) $ we need the following lemma.

\begin{Lemma}
\label{lem01}
For each $n \in \mathbb{N} $
\begin{equation}
\label{b11}
\sum_{\begin{array}{c}
k \neq \pm n \\
k \in n + 2\mathbb{Z}
\end{array}}
\frac{1}{|\lambda - k^2 |^2}  <  \frac{9}{n^2}
\quad \text{if} \quad  (n-1)^2 \leq Re \, \lambda \leq (n+1)^2,
\; \lambda \in \mathbb{C}.
\end{equation}
\end{Lemma}

\begin{proof}
Let  $k \in n + 2 \mathbb{Z} $
in all sums that appear in the proof.
The sum in (\ref{b11})
does not exceed $2 S_1 (n) + 2 S_2 (n), $
where
\begin{equation}
\label{b12}
S_1 (n) = \sum_{0\leq k < n-1} \frac{1}{[(n-1)^2 - k^2]^2}, \quad
S_2 (n) = \sum_{k > n+1} \frac{1}{[k^2 -(n+1)^2]^2}.
\end{equation}
Obviously,
\begin{equation}
\label{b13}
S_1 (1) = 0, \quad S_1 (2) = 1, \quad S_1 (3) = 1/9.
\end{equation}
If $ n \geq 4, $ then, by the inequality
$$
(n-1)^2 - k^2 = (n-1-k)(n-1+k ) \geq (n-1-k)(n-1), \quad 0 \leq k < n-1,
$$
we have
\begin{equation}
\label{b14}
S_1 (n) \leq \frac{1}{(n-1)^2} \sum_{0 \leq k <n-1}
\frac{1}{(n-1-k)^2} <  \frac{1}{(n-1)^2} \cdot \frac{\pi^2}{8}
< \frac{2}{n^2} \cdot \frac{\pi^2}{8}, \;\; n \geq 4.
\end{equation}
Also,
$$
k^2 - (n+1)^2 = (k+n+1)(k - n-1) > 2n (k-n-1)  \quad \text{if} \quad k > n+1,
$$
and therefore,
\begin{equation}
\label{b15}
S_2 (n) \leq \frac{1}{4n^2} \sum_{k> n+1}
\frac{1}{(k-n-1)^2} =
\frac{1}{4n^2} \cdot \frac{\pi^2}{8}.
\end{equation}
Now, (\ref{b14}) and (\ref{b15})
yield  that 
the sum (\ref{b11}) does not exceed $ 9 \pi^2/(16n^2) < 6/n^2 $
for $ n \geq 4. $
Since $ \pi^2 \leq 10,$ we obtain, by (\ref{b15}), that
$$
S_2 (1) \leq  \frac{\pi^2}{32} < \frac{1}{3}, \quad
S_2 (2) \leq  \frac{\pi^2}{128} < \frac{1}{12}, \quad
S_2 (3) \leq  \frac{\pi^2}{288} < \frac{1}{28};
$$
thus,
in view of (\ref{b13}),
the inequalities
(\ref{b11}) hold for each $n \in \mathbb{N}.$
\end{proof}

Let
\begin{equation}
\label{b16}
H_0 = \{ z \in \mathbb{C}: \;\; Re \, z \leq 1 \}, \qquad
H_1 = \{ z \in \mathbb{C}: \;\; Re \, z \leq 4 \},
\end{equation}
\begin{equation}
\label{b17}
H_n = \{ z \in \mathbb{C}: \;\;(n-1)^2 \leq Re \, z \leq (n+1)^2 \},
\quad n \geq 2,
\end{equation}
and
\begin{equation}
\label{b18}
D_n (r) = \{ z \in \mathbb{C}: \;\; |z - n^2 | < r \}, \quad r>0, \;
n \in \mathbb{Z}_+.
\end{equation}
Next, we estimate the norm of $ R_{\lambda}^0 V,  $
(or, $A(\lambda)$, see (\ref{b10}))
for  $\lambda \in \mathbb{C} \setminus D^+ $ if $bc = Per^+, $ and for
$\lambda \in \mathbb{C} \setminus D^- $ if $bc = Per^-, $
where
\begin{equation}
\label{b19}
D^+ = \bigcup_{k \in 2\mathbb{N}}  D_k (r), \qquad
D^- = \bigcup_{k \in 2\mathbb{N}-1}  D_k (r), \quad r= 4\|v\|.
\end{equation}
By (\ref{b16}) and (\ref{b17}),
$$ \mathbb{C} = \bigcup_{k \in 2\mathbb{N}} H_k =
\bigcup_{k \in 2\mathbb{N}-1}  H_k,
$$
and therefore, in view of (\ref{b19}),
$$
\mathbb{C} \setminus D^+ = \bigcup_{k \in 2\mathbb{N}}
\left ( H_k \setminus  D_k (r) \right ), \qquad
\mathbb{C} \setminus D^- = \bigcup_{k \in 2\mathbb{N}-1}
\left ( H_k \setminus  D_k (r) \right ).
$$
If $\lambda \in H_n \setminus  D_n (r), $  
 $ n \geq 2, $ then
(\ref{b11}) from Lemma \ref{lem01} yields 
\begin{equation}
\label{b21}
A(\lambda) \leq \frac{2}{r^2} + 9, \qquad
\lambda \in H_n \setminus  D_n (r), \;\; n \geq 2.  
\end{equation}
If $ \lambda \in H_1, $
then
$ \sup \{ A(\lambda): \; Re \, \lambda \leq 0 \}  \leq  A(0); $
thus, by (\ref{b11}) with $n=1,$
we have  
\begin{equation}
\label{b22}
A(\lambda) \leq \frac{2}{r^2} + 9, \qquad
\lambda \in H_1 \setminus  D_1 (r). 
\end{equation}
If $ \lambda \in H_0, $
i.e., $ Re \, \lambda \leq 1, $ 
then
$$
\sum_{k \in 2\mathbb{N}}  \frac{1}{|\lambda - k^2|^2}
\leq \sum_{k \in 2\mathbb{N}}  \frac{1}{|k^2-1|^2}
\leq \frac{1}{9}
\sum_{k \in 2\mathbb{N}}  \frac{1}{(k-1)^2} = \frac{\pi^2}{72}
\leq \frac{1}{7}
$$
(because $ k^2 -1 = (k+1)(k-1) \geq 3(k-1) $).
Thus,
\begin{equation}
\label{b23}
A(\lambda) \leq \frac{1}{r^2} + 2, \qquad
\lambda \in H_0 \setminus  D_0 (r). 
\end{equation}

Now, in view of (\ref{b10})
and (\ref{b21})--(\ref{b23}),
we have, with $r = 4 \|v\| \leq 1, $
that
$$
\|R^0_\lambda  V\|^2 \leq A(\lambda ) \|v\|^2 \leq 
2/16 + 9/16  \leq \frac{3}{4}, $$
for  $\lambda \in \mathbb{C} \setminus D^+ $ if $bc = Per^+, $ and for
$\lambda \in \mathbb{C} \setminus D^- $ if $bc = Per^-, $
respectively.
Therefore, $R_\lambda (L_{Per^+}) $ is well defined
for  $\lambda \in \mathbb{C} \setminus D^+, $
and $R_\lambda (L_{Per^-}) $ is well defined
for $\lambda \in \mathbb{C} \setminus D^- ,$
i.e.,
$$
\sigma (L_{Per^+}) \subset D^+ \quad \text{and} \quad
\sigma (L_{Per^-}) \subset D^-
$$
for all potentials $v $ such that $\|v\| \leq 1/4, $
and in particular for any $v_\tau (x) = \tau  v(x), \; \tau \in [0,1]. $
For each even $k$ (if $ bc = Per^+$), and for each odd $k$  
(if $ bc = Per^-$), the resolvents $R_\lambda (L_{v_\tau}) $
are analytic in $\lambda $ for
$ \lambda \in H_k \setminus D_k (4\|v\|), $
and continuous in $\tau $ for $ \tau \in [0,1].$
Thus, $\dim P_k (\tau), $ where
$$
P_k (\tau) = \frac{1}{2 \pi i} \int_{|z-k^2|= r}  (z-L_\tau)^{-1} dz,
\qquad  r= 4\|v\|, 
$$
being an integer, is a constant, i.e.,
$$
\dim P_k (1) = \dim P_k (0) =2 \quad \text{for} \;\; k >0, \quad
\dim P_0 (1) = \dim P_0 (0) =1.
$$
Hence, if $k> 0$ then the disc $D_k (4\|v\|) $ contains exactly {\em two}
(periodic, if $k$ is even, and 
antiperiodic, if $k> 0 $ is odd) eigenvalues,  
and the disc $D_0 (4\|v\|) $ contains exactly
{\em one} periodic eigenvalue. In view of (\ref{b3}) the latter proves
Part (a), i.e., (\ref{b4}) holds.\vspace{3mm}

(b) Next we prove Part (b), (\ref{b5}), by using the same
notations, but with $r=1,$ and almost the
same argument that
has been used to prove Part (a).
Thus, (\ref{b5}) will be proven if we explain that
for large enough $n$
the resolvent $R_\lambda $ is defined
for $ \lambda \in H_n \setminus D_n (1). $

With $T= R^0_\lambda V $ we have, in view of (\ref{b8}), that
$$
R_\lambda = (1-T)^{-1}  R^0_\lambda =
(1-T^2)^{-1} (1+T)  R^0_\lambda
$$
is well defined if $\|T^2 \| < 1. $
Thus, our goal is
to show that
there exists $N_0 $ such that if $n \geq N_0 $  then
$\|T^2 \| < 1 $ for 
$ \lambda \in H_n \setminus D_n (1). $

In view of (\ref{b9}), the matrix representation of $T^2 $ is
\begin{equation}
\label{b24}
(T^2)_{km} = \sum_{s \in n + 2\mathbb{Z}}
 \frac{V(k-s)V(s-m)}
{(\lambda -k^2)(\lambda - s^2)},  \quad k,m \in n + 2 \mathbb{Z}.
\end{equation}

For $\lambda \in H_n \setminus D_n (1) $
we have $ |\lambda - n^2| \geq 1 ,$
and therefore,
\begin{equation}
\label{b25}
\sum_{k,m} |(T^2)_{km}|^2  \leq 3 \left (
\Sigma_+ + \Sigma_- + \Sigma_2      \right ),
\end{equation}
where
$$
\Sigma_\pm =\sum_{k,m}
\frac{|V(k \pm n)|^2 |V(\mp n-m)|^2}{|\lambda -k^2|^2},
\quad  
\Sigma_2 = \sum_{k,m} \left | \sum_{s \neq \pm n}  \frac{V(k-s)V(s-m)}
{(\lambda -k^2)(\lambda - s^2)} \right |^2. 
$$

Taking into account that $V(0) = 0$
we have, by 
Lemma \ref{lem01},
\begin{equation}
\label{bb26}
\Sigma_{\pm} =   \left ( \frac{|V(\pm 2n)|^2}
{|\lambda - n^2 |^2} + \sum_{k \neq \pm n}
 \frac{|V(k\pm n)|^2}{|\lambda -k^2|^2} \right ) \sum_m |V(\mp n -m)|^2
\end{equation}
$$ \leq \left ( |V(\pm 2n)|^2 + \frac{9}{n^2} \|v\|^2 \right ) \|v\|^2.
$$
On the other hand, by the Cauchy inequality and Lemma \ref{lem01},
\begin{eqnarray}
\Sigma_2 \leq \sum_k \left (
\sum_{s \neq \pm n} \frac{|V(k-s)|^2}{|\lambda - k^2|^2}
\right )
\left ( \sum_m \sum_{s \neq \pm n}
\frac{|V(s-m)|^2}{|\lambda - s^2|^2} \right ) \nonumber \\
{} \label{bb27} \\
\leq \left ( \sum_k \frac{1}{|\lambda - k^2|^2} \right )
\left ( \sum_{s \neq \pm n} \frac{1}{|\lambda - s^2|^2} \right )
\cdot \|v\|^4 \leq \left (2 + \frac{9}{n^2} \right )
\cdot \frac{9}{n^2} \cdot \|v\|^4. \nonumber
\end{eqnarray}

In view of (\ref{b25}), the estimates obtained in
(\ref{bb26}) and (\ref{bb27}) show that
the Hilbert-Schmidt norm of $T^2,$ (and therefore, the operator norm 
in $\ell^2 $
of $T^2 $) goes to 0 as $n \to \infty. $  Thus,
we may choose $N_0 $ 
so that 
$$
\|T^2 \| \leq 1/2 \qquad
\text{for} \;\; n \geq N_0, \; \lambda \in H_n \setminus D_n (1).
$$
Then, as in the proof of (a), a homotopy type argument completes the proof.
\end{proof}

2. In \cite{DM3}, Theorem 8, we obtained an asymptotic formula for
the spectral gaps
$\gamma_n = \lambda^+_n - \lambda^-_n  $
as $n \to \infty. $
In this section we explain that, in fact,
the same formula gives the asymptotics of
$\gamma_n $  for each fixed $n,$
if we consider $\gamma_n $
as a function of $v$ and look for its asymptotic
as $\|v\|\to  0.$

Suppose that $ \lambda = n^2 +z, \; n \in \mathbb{N} $ is
a periodic  (or antiperiodic) eigenvalue of $L,$
with $|z| < 1, $ and $ y \neq 0 $ is a corresponding eigenfunction.
Let $ E^0_n = [e_{-n} , e_n ] $
be the eigenspace of $L^0 $ corresponding to $n^2 ,$
and let $P^0_n $ be the orthogonal projector on $E^0_n .$
We set 
$$Q^0_n = 1- P^0_n  . $$ 
Then the equation $(n^2 +z -L)y = 0 $ is equivalent to 
the following system of two equations: 
\begin{equation}
\label{2.1}
Q^0_n (n^2 +z -L^0 - V) Q^0_n y
+ Q^0_n (n^2 +z -L^0 - V) P^0_n y = 0,
\end{equation}
\begin{equation}
\label{2.2}
P^0_n (n^2 +z -L^0 - V) Q^0_n y
+ P^0_n (n^2 +z -L^0 - V) P^0_n y = 0.
\end{equation}
Taking into account that
$$ P^0_n Q^0_n = Q^0_n P^0_n = 0, \quad
P^0_n L^0 Q^0_n = Q^0_n L^0 P^0_n = 0, \quad
L^0 P^0_n y = n^2 P^0_n y,
$$
we obtain that (\ref{2.1}) and (\ref{2.2})
can be rewritten as 
\begin{equation}
\label{2.3}
Q^0_n (n^2 +z -L^0 - V) Q^0_n y
- Q^0_n V P^0_n y = 0,
\end{equation}
\begin{equation}
\label{2.4}
- P^0_n V Q^0_n y
- P^0_n V P^0_n y  + z P^0_n y = 0
\end{equation}

Let $\mathbb{H}(n) $ denote the range 
of the operator $Q^0_n . $

The operator
\begin{equation}
\label{2.5}
  A =  A(n,z):=  Q^0_n (n^2  + z - L^0 - V) Q^0_n : \;\;
\mathbb{H}(n) \to \mathbb{H}(n) 
\end{equation}
is invertible if $3\|v\|/n < 1 $
(see below Lemma \ref{lem03}).
Thus, solving (\ref{2.3}) for $ Q^0_n y $ , we obtain
$$ Q^0_n y = A^{-1} Q^0_n V P^0_n y ,$$
where $P^0_n y \neq 0 $ (otherwise
$Q^0_n y = 0, $ which implies $ y= P^0_n y + Q^0_n y =0).$
Therefore (\ref{2.4}) implies
\begin{equation}
\label{2.6}
[P^0_n V A^{-1}Q^0_n V P^0_n  + P^0_n V P^0_n - z ] P^0_n y = 0
\end{equation}
with $P^0_n y \neq 0 .$
Let
$$
\begin{pmatrix}
S^{11} & S^{12} \\
S^{21}  &  S^{22}
\end{pmatrix}
$$
be the matrix representation
of the two-dimensional operator
\begin{equation}
\label{2.7}
  S:= P^0_n V A^{-1}Q^0_n V P^0_n  + P^0_n V P^0_n : \;\;
E^0_n \to E^0_n  
\end{equation}
with respect to the basis $e_{-n} , e_n .$
Then we have
\begin{equation}
\label{2.8}
S^{11} = \langle S e_{-n} , e_{-n} \rangle ,\;
S^{22} = \langle Se_n , e_n \rangle, \;
S^{21} = \langle Se_{-n} , e_n \rangle , \;
S^{12} = \langle Se_n , e_{-n} \rangle .
\end{equation}
Since $P_n^0 y \neq 0, $ 
(\ref{2.6}) implies
\begin{equation}
\label{2.10}
\left |
\begin{array}{cc}
S^{11} - z & S^{12}\\
S^{21}  &  S^{22} - z
\end{array}
\right| = 0.
\end{equation}

In the selfadjoint case (where $v$ is real-valuead),
if $\lambda $ is a double eigenvalue, then there exists another
eigenvector $\tilde{y} $ (corresponding to $\lambda $),
such that $y $ and $\tilde{y} $ are linearly independent.
Then $P_n^0 y$ and $P_n^0 \tilde{y} $ are linearly independent also.
Indeed, if $P_n^0 y  = c P_n^0 \tilde{y} $ then
$$ Q_n^0 y = A^{-1} Q_n^0 V P_n^0 y =
c A^{-1} Q_n^0 V P_n^0 \tilde{y} = c Q_n^0 \tilde{y},
$$
which leads to a contradiction:
$$ y = P_n^0 y + Q_n^0 y = c \left ( P_n^0 \tilde{y} + Q_n^0 \tilde{y}
\right ) = c \tilde{y}.
$$
Thus $S \equiv 0 ,$ i.e., if $ \lambda = \pi n + z $ is a double eigenvalue
of a self-adjoint Schr\"odinger operator $L,$ then
(for large enough $n$)
\begin{equation}
\label{2.11}
S^{11} - z= 0, \quad  S^{12} = 0, \quad
S^{21}=0, \quad  S^{22} - z = 0. 
\end{equation}

Next, in order to obtain explicit formulas for $S^{11}, S^{22}, S^{12} $ and
$S^{21}, $
we compute the matrix representations
of the operators $A$ and $S$ with respect to the basis
$\{ e_{2k}, \; k\in \mathbb{Z} \}$ for even $n,$
and with respect to the basis
$\{ e_{2k-1}, \; k\in \mathbb{Z} \} $ for odd $n.$
The operator
$$
Q^0_n (n^2 + z - L^0 ) Q^0_n  : \; H(n) \to H(n)
$$
is invertible for any $z$ with $|z| < 1. $
Let $D_n $ denote its inverse operator.
Obviously, the matrix representing $D_n $ is
\begin{equation}
\label{2.12}
(D_n)_{km} = \frac{1}{n^2 - k^2 +z} \delta_{km},
\qquad  k,m \in (n+2\mathbb{Z})\setminus \{\pm n\},
\end{equation}
where $ \delta_{km} = 0 $ for $k\neq m$  and
$ \delta_{km} = 1$ for $k = m.$

The operator $A$ from (\ref{2.5}) can be written as
$$ A = \left [ Q^0_n (n^2 + z - L^0 ) Q^0_n \right ] (1 - T_n ), $$
where
\begin{equation}
\label{2.13}
T_n =  D_n Q^0_n V Q^0_n .
\end{equation}
Thus $A = A(n,z) $ is invertible if and only if
$1- T_n $ is invertible, and in this case
\begin{equation}
\label{2.14}
A^{-1} = (1- T_n )^{-1} D_n .
\end{equation}

The matrix representation
of the operator of multiplication by
$v(x)= \sum V(k) \exp (ikx ) $ is
\begin{equation}
\label{2.15}
V_{km} = V(k-m).
\end{equation}
Now, by (\ref{2.12}), (\ref{2.13}) and (\ref{2.15})
we obtain
that the matrix representation of the operator $T_n $ is
given by
\begin{equation}
\label{2.16}
(T_n)_{km} = \frac{V(k-m)}{n^2 -k^2 +z}, \qquad
k,m \in (n+2\mathbb{Z})\setminus \{\pm n\}.
\end{equation}

\begin{Lemma}
\label{lem03}
If $ |z| \leq 1,$
then, for each $n\in \mathbb{N}, $
the norm of the operator $T_n : H_n \to H_n $ satisfies
\begin{equation}
\label{2.17}
\|T_n \| \leq  \frac{3\|v\|}{n}.
\end{equation}
\end{Lemma}

\begin{proof}
Since the $\ell^2 $-norm does not exceed the Hilbert-Schmidt
norm, we obtain, in view of (\ref{2.16}) and Lemma \ref{lem01}, 
$$ \|T_n \|^2 \leq 
\sum_k
\frac{1}{|n^2 -k^2 +z|^2} \sum_m |V(k-m)|^2
\leq \frac{9\|v\|^2}{n^2},
$$
where $k,m \in (n+2\mathbb{Z})\setminus \{\pm n\}. $
\end{proof}

Let us consider $n \geq 9 \|v\|$ until the end of this section.
Then, 
by (\ref{2.17}),
$ \|T_n \| \leq 3\|v\|/n \leq 1/3, $ so 
(\ref{2.14}) yields
\begin{equation}
\label{2.18}
A^{-1} = \sum_{m=0}^\infty T_n^m D_n,
\end{equation}
and therefore, by (\ref{2.7}),  
\begin{equation}
\label{2.19}
S = P_n^0 V P_n^0 + \sum_{m=0}^\infty  P_n^0 V T_n^m D_n Q_n^0 V P_n^0.
\end{equation}
Now, by (\ref{2.12}), (\ref{2.15}) and (\ref{2.16}) we obtain,
in view of (\ref{2.8}), that
\begin{equation}
\label{2.20}
S^{ij} (n, z) = \sum_{k=0}^\infty S^{ij}_k (n,z), \quad i,j =1,2,
\end{equation}
where
\begin{equation}
\label{2.21}
S^{11}_0 = S^{22}_0 = 0, \quad S^{12}_0 = V(-2n), \quad
S^{21}_0 = V(2n), 
\end{equation}
and for each $k=1,2, \ldots, $
\begin{equation}
\label{2.22}
S_k^{11} (n,z) =  \sum_{j_1, \ldots, j_k \neq \pm n}
\frac{V(-n-j_1)V(j_1 - j_2) \cdots V(j_{k-1} -j_k) V(j_k +n)}
{(n^2 -j_1^2 +z) \cdots (n^2 - j_k^2 +z)}
\end{equation}
\begin{equation}
\label{2.23}
S_k^{22} (n,z) =  \sum_{j_1, \ldots, j_k \neq \pm n}
\frac{V(n-j_1)V(j_1 - j_2) \cdots V(j_{k-1} -j_k) V(j_k -n)}
{(n^2 -j_1^2 +z) \cdots (n^2 - j_k^2 +z)}
\end{equation}
\begin{equation}
\label{2.24}
S_k^{12} (n,z) = 
\sum_{j_1, \ldots, j_k \neq \pm n}
\frac{V(-n-j_1)V(j_1 - j_2) \cdots V(j_{k-1} -j_k) V(j_k -n)}
{(n^2 -j_1^2 +z) \cdots (n^2 - j_k^2 +z)}
\end{equation}
\begin{equation}
\label{2.25}
S_k^{21} (n,z) =  
\sum_{j_1, \ldots, j_k \neq \pm n}
\frac{V(n-j_1)V(j_1 - j_2) \cdots V(j_{k-1} -j_k) V(j_k +n)}
{(n^2 -j_1^2 +z) \cdots (n^2 - j_k^2 +z)}
\end{equation}
The above series converge absolutely and uniformly for $ |z| \leq 1. $

\begin{Lemma}
\label{lem04}
(a) For any (even complex-valued) potential $v$
\begin{equation}
\label{2.26}
S^{11} (n,z) = S^{22} (n,z).
\end{equation}

(b) If $v$ is a real-valued potential, then
\begin{equation}
\label{2.27}
S^{12} (n,z) = 
\overline{S^{21} (n,\overline{z})}.
\end{equation}
\end{Lemma}

\begin{proof}
(a) By (\ref{2.22}) and (\ref{2.23}),
for each $k =1,2, \ldots, $ the change of sumation indices
$$ i_s = - j_{k+1-s}, \quad s=1, \ldots, k,
$$
proves that $S^{11}_k (n,z) = S^{22}_k (n,z). $
Thus, in view of (\ref{2.20}) and (\ref{2.21}), (\ref{2.26}) holds.

(b) If $v$ is real-valued, we have for its Fourier coefficients 
the identity $ V(-m) = \overline{V(m)}. $
By (\ref{2.20}),
$$
 S^{12}_0 (n,z) = V(-2n) = \overline{V(2n)} =
\overline{S^{21}_0 (n, \overline{z})}. $$
Also, for each $ k =1,2, \ldots, $
the change of sumation indices
$$ i_s = j_{k+1-s}, \quad s=1, \ldots, k, $$
explains that 
$ S^{12}_k (n,z) = 
\overline{S^{21}_k (n,\overline{z})}, $
thus (\ref{2.27}) holds.
\end{proof}

In this paper we consider only real-valued potentials $v.$
For convenience we set
\begin{equation}
\label{2.31}
\alpha_n (z) = S^{11} (n,z) = S^{22} (n,z), \quad
\beta_n (z) = S^{21} (n,z) = \overline{S^{12} (n, \overline{z})}.
\end{equation}
Under these notations the basic equation (\ref{2.10}) becomes
\begin{equation}
\label{2.32}
\left (z - \alpha_n (z) \right )^2 - |\beta_n (z) |^2 = 0,
\end{equation}
which splits into two equations
\begin{equation}
\label{2.33}
z - \alpha_n (z) -  |\beta_n (z)| = 0,
\end{equation}
\begin{equation}
\label{2.34}
z - \alpha_n (z) + |\beta_n (z)| = 0.
\end{equation}

\begin{Lemma}
\label{lem05}
If $|z| \leq 1 $ and $ \|v\|/n \leq 1/9,$ then
\begin{equation}
\label{2.35}
 \left | \frac{d}{dz}\alpha_n (z) \right | \leq \frac{\|v\|^2}{n^2},
 \qquad
 \left | \frac{d}{dz}\beta_n (z) \right | \leq \frac{\|v\|^2}{n^2}.
\end{equation}
\end{Lemma}

\begin{proof}
By (\ref{2.5}) and (\ref{2.7})
$$
\frac{d}{dz} S(n;z) = - P_n^0 V Q_n^0 (A^{-1})^2 Q_n^0 V  P_n^0,
$$
and therefore,
in view of (\ref{2.8}) and (\ref{2.31}), we have
\begin{eqnarray}
\label{2.36}
\alpha^\prime_n (z)  = -
\langle P_n^0 V Q_n^0 (A^{-1})^2 Q_n^0 V  P_n^0 e_n, e_n \rangle,\\
\beta_n^\prime (z) = -
\langle P_n^0 V Q_n^0 (A^{-1})^2 Q_n^0 V  P_n^0 e_{-n}, e_n \rangle.
\label{2.36a}
\end{eqnarray}
Set
\begin{equation}
\label{2.37}
f_{\pm n} = Q^0_n V P^0_n e_{\pm n}, \quad
h_n (x) = \langle P^0_n V Q^0_n x, e_n \rangle.
\end{equation}
With these notations we have, by  (\ref{2.36}) and (\ref{2.36a}),
\begin{equation}
\label{2.38}
\alpha^\prime_n (z)  = - h_n \left [ (A^{-1}(z))^2 f_n \right ], \qquad
\beta^\prime_n (z)  =- h_n \left [ (A^{-1}(z))^2 f_{-n} \right ], 
\end{equation}
and therefore,
\begin{equation}
\label{2.39}
|\alpha^\prime_n (z)| \leq
\| h_n \|\cdot \|A^{-1} \|^2 \cdot \| f_n \|,  \quad
|\beta^\prime_n (z)|   \leq \| h_n \| \cdot \|A^{-1}\|^2 \cdot \|f_{-n} \|. 
\end{equation}
By Lemma \ref{lem03} and (\ref{2.18}),
if $ \|v\|/n < 1/9 $ then
\begin{equation}
\label{2.40}
\|A^{-1} \| \leq \left ( \sum_{k=0}^\infty \|T_n \|^k
\right ) \cdot \|D_n \| < \frac{3}{2} \|D_n \|.
\end{equation}
The operator $D_n $ is diagonal,
and therefore,
by (\ref{2.12}),
\begin{equation}
\label{2.41}
\|D_n \| = \max_{k \neq \pm n} \frac{1}{| n^2 - k^2 +z|} .
\end{equation}
If $ n=1,$ then (\ref{2.41}) implies
\begin{equation}
\label{2.42}
\|D_1 \| = \max_{k \neq \pm 1} \frac{1}{| k^2 - 1- z| }  \leq \frac{1}{7}.
\end{equation}
For $ n \geq 2 $ we obtain, by (\ref{2.41}),
\begin{equation}
\label{2.43}
\|D_n \| \leq \frac{1}{4n-5} \leq \frac{2}{3n}.
\end{equation}
Now (\ref{2.40}), (\ref{2.42}) and (\ref{2.43}) yield
\begin{equation}
\label{2.44}
\|A^{-1}\| \leq 1/n \quad \text{if} \;\; \|v\|/n < 1/9
 \;\; \text{and} \;\; |z| \leq 1.
\end{equation}

On the other hand, by (\ref{2.15}),
\begin{equation}
\label{2.45}
\|f_{\pm n}\| = \left \|  \sum_{k \neq \pm n} V(k \mp n) \, e_k 
\right \| \leq \|v\|,
\end{equation}
and, with $ x = \sum x_k e_k ,$
$$
\|h_n (x) \| = \left | \sum_{k \neq \pm n} V(n-k) x_k \right |
\leq \|v\| \cdot \|x\|, 
$$
thus
\begin{equation}
\label{2.46}
\|h_n \| \leq \|v \|.
\end{equation}
Now (\ref{2.44})-(\ref{2.46}) yield, in view of (\ref{2.39}), that
(\ref{2.35}) holds.
\end{proof}

\begin{Theorem}
\label{thm1}
(a)  If $ \|v\| \leq 1/9, $ then for each $n=1,2,\ldots $
there exists $z = z_n $ such that
\begin{equation}
\label{d20}
|z| \leq 4\|v\|,
\end{equation}
and
\begin{equation}
\label{d21}
2|\beta_n (z)| \left (1 -3 \|v\|^2/n^2 \right ) \leq
\gamma_n \leq 2|\beta_n (z)| \left (1 + 3 \|v\|^2/n^2 \right ).
\end{equation}

(b) If $V(0) = \frac{1}{\pi} \int_0^\pi v(x) dx = 0, $ then
there is $N_0 = N_0 (v) $ such that
(\ref{d21}) holds
for $ n \geq N_0 $
with $z = z_n, $ 
\begin{equation}
\label{d22}
|z_n| < 1. 
\end{equation}
\end{Theorem}

\begin{proof}
We will prove (\ref{d21}) simultaneously in both cases (a) and (b).

In Case (a) we know,
by part (a) of Proposition \ref{prop1}, that, for each $n\in \mathbb{N},$
there are exactly two eigenvalues $\lambda^\pm_n =  n^2 + z^\pm_n $ of $L$
(periodic for even $n$ and antiperiodic for odd $n$)
such that $ |z^\pm_n | < 4\|v\|< 1. $

By part (b) of Proposition \ref{prop1},
the same is true in Case (b) if $n$ is large enough.
Let $N_0 =N_0 (v) $ be chosen so that part (b) of Proposition \ref{prop1}
holds for $ n \geq N_0, $ and
\begin{equation}
\label{d23}
\|v\|/n  \leq 1/9 \qquad \text{for} \;\;  n \geq N_0.
\end{equation}
Fix an $ n\geq N_0 $ in Case (b), and  let $n\in \mathbb{N} $
in Case (a).

We know, in both cases (a) and (b), that
the numbers $z^\pm_n $
are roots of (\ref{2.10}), and therefore, of (\ref{2.32}).
If $z^+_n = z^-_n = z^*, $
then $\lambda = \pi n + z^* $ is a double eigenvalue and
(\ref{2.11}) yields
$\beta_n (z^* ) = 0,$ thus (\ref{d21}) holds.

If
$z^+_n \neq  z^-_n, $
set
\begin{equation}
\label{d37}
\zeta^+_n = z^+_n - \alpha_n (z^+_n), \qquad
\zeta^-_n = z^-_n - \alpha_n (z^-_n).
\end{equation}
Then, by (\ref{2.33}) and (\ref{2.34}),
\begin{equation}
\label{d38}
|\zeta^+_n | = |\beta_n (z^+_n)|, \qquad
|\zeta^-_n | = |\beta_n (z^-_n)|.
\end{equation}
By (\ref{d37}),
$$ \zeta_n^+ - \zeta_n^+ = \int_{z^-_n}^{z^+_n}
\left ( 1- \alpha_n^\prime (z) \right ) dz.
$$
Thus, in view of
Lemma \ref{lem05}, and (\ref{d23}) in Case (b),
or the inequality $\|v\| \leq 1/9 $ in Case (a), we obtain 
\begin{equation}
\label{d40}
 (z^+_n - z^-_n ) (1 - \|v\|^2/n^2) \leq 
|\zeta^+_n - \zeta^-_n | \leq (z^+_n - z^-_n ) (1 + \|v\|^2/n^2),
\end{equation}
which yields
\begin{equation}
\label{d41}
 |\zeta^+_n - \zeta^-_n | \left (1 - \frac{\|v\|^2}{n^2} \right ) \leq 
z^+_n - z^-_n 
\leq |\zeta^+_n - \zeta^-_n |
\left (1 + \frac{9\|v\|^2}{8n^2} \right ) \leq
\frac{9}{8} |\zeta^+_n - \zeta^-_n |.
\end{equation}

Since $ z^+_n $ and $z^-_n $ are roots of (\ref{2.32}),
each of these numbers is a root of either (\ref{2.33}), or (\ref{2.34}).
There are two cases: (i) $z^+_n $ and $z^-_n $ are roots of different
equations; (ii) $z^+_n $ and $z^-_n $ are roots of one and the same equation.

In Case (i) we have, by (\ref{2.33}), (\ref{2.34}) and (\ref{d38}), that
\begin{equation}
\label{d42}
 |\zeta^+_n - \zeta^-_n |
= |\beta_n (z^+_n)| +|\beta_n (z^-_n)| =
 |\zeta^+_n| + |\zeta^-_n |.
\end{equation}
On the other hand,
since $\beta_n (z^+_n) - \beta_n (z^-_n)
= \int_{z^-_n}^{z^+_n} \beta^\prime_n (t) dt ,$
Lemma \ref{lem05} and (\ref{d23}) in Case (b),
or the inequality $\|v\| \leq 1/9 $ in Case (a), 
imply that
\begin{equation}
\label{d43}
|\beta_n (z^+_n) - \beta_n (z^-_n)|
\leq
(z^+_n -z^-_n ) \frac{\|v\|^2}{n^2} \leq
 |\zeta^+_n - \zeta^-_n | \frac{9\|v\|^2}{8n^2} .
\end{equation}
Thus (\ref{d38}) and (\ref{d42}) yield
$$
\left | |\zeta^+_n | - |\zeta^-_n| \right | =
\left |  |\beta_n (z^+_n)| - |\beta_n (z^-_n)| \right |
\leq
\left ( |\zeta^+_n | + |\zeta^-_n| \right ) \frac{9\|v\|^2}{8n^2}.
$$
so, since $2 |\zeta^+_n |=
\left ( |\zeta^+_n | + |\zeta^-_n| \right ) +
\left ( |\zeta^+_n | - |\zeta^-_n| \right ), $
$$
\left ( |\zeta^+_n | + |\zeta^-_n| \right )
\left (1 - \frac{9\|v\|^2}{8n^2} \right )
\leq 2|\zeta^+_n |  \leq 
\left ( |\zeta^+_n | + |\zeta^-_n| \right )
\left (1 + \frac{9\|v\|^2}{8n^2} \right ),
$$
and therefore, since $\|v\|^2/n^2 <1/9,$
\begin{equation}
\label{d44}
2|\zeta^+_n |  \left (1 - \frac{9\|v\|^2}{8n^2} \right )
\leq |\zeta^+_n | +  |\zeta^-_n | \leq
2|\zeta^+_n |  \left (1 + \frac{8\|v\|^2}{7n^2} \right )
\end{equation}
Since $ \gamma_n = z^+_n - z^-_n ,$ 
(\ref{d41}), (\ref{d42}) and (\ref{d44}) yield (\ref{d21})
with $ z = z_n^+.$

Case (ii), where $z^+_n $ and $z^-_n $ are simultaneously roots of
one of the equations (\ref{2.33}) and (\ref{2.34}), is impossible.
Indeed, by (\ref{d43}) we would have (since $\|v\|^2/n^2 <1/9)$
$$
 |\zeta^+_n - \zeta^-_n | =
\left |  |\beta_n (z^+_n)| - |\beta_n (z^-_n)| \right |
\leq
|\zeta^+_n - \zeta^-_n | \cdot \frac{9\|v\|^2}{8n^2} \leq
\frac{1}{72} |\zeta^+_n - \zeta^-_n |,
$$
which implies $\zeta^+_n = \zeta^-_n .$
But then (\ref{d41}) yield $z^+_n = z^-_n ,$
which is a contradiction to our assumption that
$z^+_n \neq z^-_n .$
\end{proof}

\section{Asymptotic formula for
the spectral gaps of a Schr\"odinger 
operator with a two term potential}

In this section we apply the general asymptotic formula (\ref{d21})
from Theorem~\ref{thm1}
to get a corresponding formula for
a Hill operator with a potential of the form
$$v(x) = a \cos 2x + b \cos 4x, \qquad
a= -4\alpha t, \quad b = - 2 \alpha^2.
$$
The next theorem gives the asymptotics of $\gamma_n = \gamma_n (\alpha, t), $
for fixed $n$ and $t,$  as $\alpha \to 0.$

\begin{Theorem}
\label{thm2}
Let $\gamma_n, \, n \in \mathbb{N} $
be the lengths of instability zones of
the Hill operator
\begin{equation}
\label{3.0}
Ly = - y^{\prime \prime} - [4\alpha t \cos 2x + 2\alpha^2 \cos 4x ]y,
\end{equation}
where either both $\alpha$ and $ t $ are real, or both are pure
imaginary numbers.

If $t$ is fixed and $ \alpha \to 0, $ then for even $n$ 
\begin{equation}
\label{3.1}
\gamma_n = \left | \frac{8\alpha^n}{2^n [(n-1)!]^2} \prod_{k=1}^{n/2}
\left ( t^2 - (2k-1)^2 \right ) \right |
\left ( 1 + O(\alpha) \right ),
\end{equation}
and for odd $n$
\begin{equation}
\label{3.2}
\gamma_n = \left | \frac{ 8\alpha^n t}{2^n [(n-1)!]^2}
\prod_{k=1}^{(n-1)/2}
\left ( t^2 - (2k)^2 \right ) \right | \left ( 1 + O(\alpha ) \right ).
\end{equation}
\end{Theorem}

\begin{proof} For convenience the proof is divided into several steps.

{\em Step 1.}  First we apply (for small enough $\alpha $)
Theorem \ref{thm1} to the Hill operator with the potential
$v(x) = - 4\alpha t \cos 2x - \alpha^2 \cos 4x, $ i.e.,
with
\begin{equation}
\label{3.3}
V(\pm 2)= -2t \alpha, \quad V(\pm 4) = - \alpha^2, \quad
V(k) = 0  \; \mbox{if} \; k\neq \pm 2, \pm 4.
\end{equation}

Since
\begin{equation}
\label{3.3a}
 \|v\|^2 = 8 |t|^2 |\alpha|^2 + 2 |\alpha|^4
\end{equation}
we obtain, by (\ref{d20}) and (\ref{d21}), that
\begin{equation}
\label{3.4}
\gamma_n =\pm 2 \left ( V(2n) +  \sum_{k=1}^\infty \beta_k (n,z) \right )
(1+ O(|\alpha|^2 ),
\end{equation}
where $ z= z(n), $
\begin{equation}
\label{3.4a}
  z = O(\alpha),
\end{equation}
\begin{equation}
\label{3.5}
\beta_k (n,z) = \sum_{j_1, \ldots j_k \neq \pm n}
\frac{V(n+j_1 ) V(j_2 - j_1 ) \cdots V(n-j_k)}
{(n^2 - j_1^2 + z) \cdots (n^2 - j_k^2 + z)}, 
\end{equation}
and all series converge absolutely and uniformly for small enough $\alpha.$

Observe that each non-zero term in (\ref{3.5}) corresponds to
a $k$-tuple of indices $(j_1, \ldots, j_k)$ such that
\begin{equation}
\label{3.6}
(n+j_1)+(j_2 - j_1) + \cdots + (j_k - j_{k-1}) + (n-j_k) = 2n
\end{equation}
and, by (\ref{3.3}),
\begin{equation}
\label{3.7}
(n+j_1), (j_2 - j_1), \ldots, (j_k - j_{k-1}), (n-j_k) \in
\{\pm 2, \pm 4 \}.
\end{equation}
Therefore, in view of (\ref{3.6}) and (\ref{3.7}),
there is one-to-one correspondence between
the non-zero terms of (\ref{3.5}) and the walks from $-n$ to $n$
with steps $\pm 2$ and $\pm 4.$

By (\ref{3.3}), each non-zero expression of the form
\begin{equation}
\label{3.8}
V(n+j_1 ) V(j_2 - j_1 ) \cdots V(n-j_k)
\end{equation}
is a monomial in $\alpha $ 
of degree
$$ \frac{1}{2} \left (|n+j_1| + |j_2 -j_1 | + \cdots + |n-j_k |
\right ) . $$
Therefore, $n $ is the minimal possible degree, and
each such monomial of degree $n$
corresponds to a walk from
$-n$ to $n$ with positive steps.
In addition, each expression of the form (\ref{3.8})
is also a monomial in $t.$

Taking into account the above remark, we obtain from
(\ref{3.4}) - (\ref{3.5}) that
\begin{equation}
\label{3.10}
\gamma_n  =\pm  P_n (t)\alpha^{n} + O \left (\alpha^{n+1} \right ).
\end{equation}

Our next goal is to find the polynomial $P_n (t) $
for each $n \in \mathbb{N}.$
In view of (\ref{3.4}) -- (\ref{3.5}),
the above discussion shows that
\begin{equation}
\label{3.11}
P_n (t) \alpha^n = 2V(2n) + 2 \sum_{-n <j_1 <\cdots <j_s <n}
\frac{V(n+j_1 ) V(j_2 - j_1 ) \cdots V(n-j_s)}
{(n^2 - j_1^2) \cdots (n^2 - j_s^2)} ,
\end{equation}
where each non-zero term corresponds to a walk with positive
steps. Moreover, the coefficient
in front of $t^n $ is
coming from the term corresponding to a walk
with steps of length 2 only. Thus we have
$P_n(t)= C_n t^n + \cdots, $ where
\begin{equation}
\label{3.12}
C_n = 2 (-2)^n \left ( \prod_{j=1}^{n-1} \left (n^2  - (-n+2j)^2 \right )
\right )^{-1} =
\frac{8(-1)^n}{2^n ((n-1)!)^2} .
\end{equation}

For even $n,$ each walk from $-n$ to $n$ has
even number of steps with length 2.
Thus, by (\ref{3.11}), we obtain that
$P_n (t) $ is a sum of monomials of even degrees, so
for $n=2m$ we have
$$ P_{2m} (t) = C_{2m}  \prod_{k=1}^m (t^2 - x_k),
$$
where $x_k, \; k=1, \ldots, m,$ depend on $m.$

For $n$ odd, say $ n= 2m-1,$
each walk from $-n$ to $n$ has odd number
of steps with length 2, and therefore, in this case
(\ref{3.11}) implies that $P_{2m-1}$ is a sum of
monomials of odd degrees. Thus we have
$$ P_{2m-1} (t) = C_{2m-1} \, t \prod_{k=1}^{m-1} (t^2 - y_k),
$$
where $y_k, \; k=1, \ldots, m-1,$ depend on $m.$

Taking into account the combinatorial meaning of the non-zero terms
in (\ref{3.11}) it is easy to compute $P_1 (t), P_2 (t), P_3 (t) $
and $P_4 (t).$ We have
\begin{equation}
\label{3.13}
P_1 (t) =- 4t, \quad
P_3 (t) = - \frac{1}{4} t(t^2 -2^2),
\end{equation}
\begin{equation}
\label{3.14}
P_2 (t) = 2(t^2 -1), \quad
P_4 (t) =\frac{1}{72} (t^2 - 1)(t^2 - 3^2).
\end{equation}
Indeed, if $n=1, $
then there is only one walk with step 2
from -1 to 1, so we obtain $P_1 (t) \alpha = 2V(2)=- 4t \alpha.$

If $n=3,$ then there are exactly three walks from -3 to 3 with positive
steps: (2,2,2), (2,4), (4,2). Now, from (\ref{3.11}) it follows that
$$ P_3 (t) \alpha^3 = \frac{-2(2t\alpha)^3}{(3^2 - (-1)^2)(3^2 - 1^2)}
+\frac{2(2t\alpha)\alpha^2}{3^2 - (-1)^2} +
\frac{2(2t\alpha)\alpha^2}{3^2 - 1^2}, $$
and therefore, the second formula in (\ref{3.13}) holds.
The formulae (\ref{3.14}) can be obtained in the same way.

Obviously, the theorem will be proved, if we show that similar
formulae hold for each $n,$ i.e.,
for $n=2m$
\begin{equation}
\label{3.15}
P_{2m} =  C_{2m} (t^2 -1) (t^2 - 3^2)\cdots (t^2 - (2m-1)^2),
\end{equation}
and
for $n=2m-1$
\begin{equation}
\label{3.16}
P_{2m-1}(t) =  C_{2m-1} t (t^2 - 2^2)\cdots (t^2 - (2m-2)^2).
\end{equation}

{\em Step 2.}  In the following,
till the end of the proof of the theorem, we assume that
both $\alpha $ and $t$ are real numbers.
Next we recall some facts from \cite{MW}, Chapter VII,
that will be used in the proof. (We discuss these constructions in detail
in \cite{DM12} where we sharpen the Magnus-Winkler results
and analyze their connections to the theory of quasi-exactly solvable
differential equations \cite{Tur96}.)

The eigenvalue equation for the operator (\ref{3.0})
is
\begin{equation}
\label{3.17}
y^{\prime \prime} +[\lambda + 4t\alpha \cos 2x + \alpha^2 \cos 4x ]y = 0.
\end{equation}
The substitution
\begin{equation}
\label{3.18}
y = u e^{\alpha \cos 2x}
\end{equation}
carries (\ref{3.17}) into the equation
\begin{equation}
\label{3.19}
u^{\prime \prime} - 4 \alpha (\sin 2x ) u^\prime
+ [ \mu  + 4(t-1) \alpha \cos 2x ]u = 0, 
\end{equation}
where $ \mu = \lambda + 2 \alpha^2 . $
If $y(x) $ is a periodic (respectively antiperiodic)
solution of (\ref{3.17}), then $u(x) =y(x) e^{-\alpha \cos 2x} $
is a periodic
(respectively antiperiodic) solution of (\ref{3.19}), and v.v.,
if $u(x) $ is a periodic (respectively antiperiodic) 
solution of (\ref{3.19}), then (\ref{3.18})
is a periodic
(respectively antiperiodic) solution of (\ref{3.17}).

Since $\sin 2x $
is an odd function
and $\cos 2x $  is an even function, it is easy to see 
that if $u(x) $ is a periodic (or antiperiodic)
solution of (\ref{3.19}), then 
the function $ \tilde{u} (x) = u(-x) $
is also a periodic (respectively antiperiodic) solution
of (\ref{3.19}).
On the other hand
$u(x) + \tilde{u} (x) $ is an even function,
and $u(x) - \tilde{u} (x) $ is odd function. Therefore,
if for some $\mu $ 
the equation (\ref{3.19}) has a non-zero solution, then 
it has also either an even non-zero solution, or 
an odd non-zero solution, or both.

In other words, when solving (\ref{3.19}), we may look for
periodic solutions of the form
\begin{equation}
\label{3.20}
u(x) = A_0 + \sum_{k \in 2\mathbb{N}} A_k \cos kx, \quad
w(x) = \sum_{k \in 2\mathbb{N}} B_k \sin kx, 
\end{equation}
or antiperiodic solutions of the form
\begin{equation}
\label{3.21}
u(x) =  \sum_{k \in 2\mathbb{N}-1} A_k \cos kx, \quad
w(x) = \sum_{k \in 2\mathbb{N}-1} B_k \sin kx.
\end{equation}

Observe that only even indices $k$ are used in (\ref{3.20}),
while only odd $k$ appear in (\ref{3.21}).

By substituting (\ref{3.20}) into (\ref{3.19})
we obtain that $u(x) $ is a periodic even solution
(respectively $v(x) $ is a periodic odd solution)
if and only if $A_k, \, k=0,2,4, \ldots, $
satisfy the recurrence relations\footnote{In \cite{MW}, p. 95,
formula (7.17) for $n=1$ (which is equivalent to (\ref{3.24}))
gives a coefficient 2 in front of $\alpha (t+1)A_0 $
although 4 is the correct coefficient.}
\begin{equation}
\label{3.23}
-\mu A_0 + 2\alpha (t-1) A_2 = 0,
\end{equation}
\begin{equation}
\label{3.24}
4 \alpha (t+1)A_0 + (2^2 - \mu ) A_2 + 2 \alpha (t-3)A_4 = 0, 
\end{equation}
\begin{equation}
\label{3.25}
 2 \alpha (t-1+k)A_{k-2} + (k^2 -\mu )A_k + 2 \alpha (t-1-k) A_{k+2} = 0,
 \quad k\geq 4
\end{equation}
and respectively $B_k, \, k=2,4, \ldots, $ satisfy
\begin{equation}
\label{3.26}
 (2^2 - \mu ) B_2 + 2 \alpha (t-3)B_4 = 0, 
\end{equation}
\begin{equation}
\label{3.27}
 2 \alpha (t-1+k)B_{k-2} + (k^2 -\mu )B_k + 2 \alpha (t-1-k) B_{k+2} = 0,
 \quad k\geq 4.
\end{equation}

If we substitute (\ref{3.21}) into (\ref{3.19}), then
it follows that $u(x) $ is an antiperiodic even solution
(respectively $w(x) $ is an antiperiodic odd solution)
if and only if $A_k, \, k=1,3,5,\ldots, $ satisfy the relations
\begin{equation}
\label{3.28}
(1-\mu + 2 \alpha t) A_1 + 2\alpha (t-2) A_3  = 0, 
\end{equation}
\begin{equation}
\label{3.29}
 2 \alpha (t-1+k)A_{k-2} + (k^2 -\mu )A_k + 2 \alpha (t-1-k) A_{k+2} = 0,
 \quad k\geq 3,
\end{equation}
and respectively $B_k, \, k=1,3,5,\ldots, $ satisfy
\begin{equation}
\label{3.30}
(1-\mu - 2 \alpha t) B_1 + 2\alpha (t-2) B_3  = 0, 
\end{equation}
\begin{equation}
\label{3.31}
 2 \alpha (t-1+k)B_{k-2} + (k^2 -\mu )B_k + 2 \alpha (t-1-k) B_{k+2} = 0,
 \quad k\geq 3.
\end{equation}

{\em Step 3.}  Now we prove (\ref{3.15}).
Observe, that to prove (\ref{3.15})
(since $ \deg P_{2m}= 2m, $ and $P_{2m} $ is even)
it is enough to show that 1 is a root of the polynomials
$P_2, P_4, \ldots, $  3 is a root of the polynomials
$P_4, P_6, \ldots, $ and so on.
So, we are going to prove 
the following statement.

{\em Claim.  For each $m=1,2, \ldots $ the number $2m-1 $
is a joint root of the polynomials}
$ P_{2m}, P_{2m+2}, \ldots.$

First we prove the claim for $m=1.$
By (\ref{3.3a}), it is easy to see that
\begin{equation}
\label{3.33}
\|v\| < 1/4   \quad \text{if} \quad t \geq 1 \quad \text{and} \quad
|\alpha| <1/(12t).
\end{equation}

Choose $t=1, $ fix an even $n\geq 2,$
and let $ |\alpha |< 1/12.$
Then, by (\ref{3.33}),
$ \|v\|< 1/4,  $ and therefore, by part (a) of Proposition~\ref{prop1},
there exists 
a periodic eigenvalue $\lambda = \lambda (\alpha ) $
of the operator
$L$ such that $|\lambda - n^2 | < 1, $
and with this $\lambda $ the equation (\ref{3.17}) 
has a non-zero solution.
Then, with $\mu = \lambda + 2 \alpha^2 \geq \lambda >0, $
the equation (\ref{3.19})
has a non-zero periodic solution.
Moreover, by Step 2, we know that
the equation (\ref{3.19})
has an even, or an odd solution
of the form (\ref{3.20}), and the corresponding coefficients
$A_k $ (respectively $B_k $) satisfy
(\ref{3.23})-(\ref{3.25})
(respectively (\ref{3.26}) and (\ref{3.27})).

If there is a non-zero periodic solution of (\ref{3.19})
of the form
$u(x) = A_0 + \sum_{k \in 2 \mathbb{N}} A_k \cos kx, $
then with $t=1$ and $\mu = \lambda + 2 \alpha^2 >0 $
we obtain, by (\ref{3.23}),
 that $A_0 = 0.$ Then
 $ w(x)= \sum_{k \in 2 \mathbb{N}} A_k \sin kx $
 is a non-zero odd solution of (\ref{3.19}), because, with
 $t=1 $ and $A_0 = 0, \; A_k = B_k, $
 the system of equations (\ref{3.24}), (\ref{3.25})
 coincides with 
 the system of equations (\ref{3.26}), (\ref{3.27}).

The same argument shows that
if there is a non-zero odd solution of (\ref{3.19})
of the form
 $ w(x)= \sum_{k \in 2 \mathbb{N}} B_k \sin kx, $
 then 
 $ u(x)= \sum_{k \in 2 \mathbb{N}} B_k \cos kx $
 is also a non-zero periodic solution.

So, in both cases the equation (\ref{3.19})
has two linearly independent periodic solutions
$u(x)$ and $w(x),$
and therefore, (\ref{3.17}) has also
two linearly independent periodic solutions
$ u(x) \exp (\alpha \cos 2x )$ and
$ w(x) \exp (\alpha \cos 2x ). $
Thus $\lambda $ is an eigenvalue of multiplicity 2,
and therefore, the $ n$-th instability zone is closed.
Hence we have for each small enough $\alpha $ that
$$ \gamma_n = \pm P_n (1) \alpha^n + O\left (\alpha^{n+1}\right ) \equiv 0,
$$
which implies that $P_n (1) = 0.$

Next we consider the general case.
Fix 
$m \geq 1, $ choose  $t= 2m +1,$
and let $|\alpha| < 1/(12t). $ 
By (\ref{3.33}), if
$\alpha < 1/(12t), $
then $ \|v \| < 1/4. $
Fix an even $n > 2m $ and choose,
by part (a) of Proposition~\ref{prop1},
a periodic eigenvalue
$\lambda = \lambda (\alpha ) $ such that $|\lambda - n^2 |< 1. $
Since $n\geq 2(m+1)$ we have
\begin{equation}
\label{3.32}
\mu = \lambda + 2 \alpha^2 > \lambda > n^2 -1 \geq 4m^2 +1.
\end{equation}

The equation (\ref{3.17}) has a non-zero periodic solution
(an eigenfunction corresponding to $\lambda$),
so the equation (\ref{3.19}), considered with
$\mu = \lambda + 2 \alpha^2,$ has also a non-zero periodic solution.
By Step 2 we know that (\ref{3.19})
has either a non-zero even periodic solution,
or a non-zero odd periodic solution.

Suppose the first case occurs. Let
 $ u(x)=A_0 + \sum_{k \in 2 \mathbb{N}} A_k \cos kx $
be a non-zero solution of (\ref{3.19}). 
Then the coefficients $A_k, k=0,2,4,\ldots, $
satisfy the system of equations (\ref{3.23})--(\ref{3.25}).

Consider the first $m+1 $ equations.
Since $ t= 2m+1  ,$ the coefficient $2\alpha (t-1-2m)$
 of $A_{2m+2} $ vanishes,
so we have $m+1 $ homogeneous equations in unknowns
$A_0, \ldots, A_m.$
The corresponding coefficient determinant $D$
has on its main diagonal $-\mu, 2^2-\mu, \ldots, (2m)^2 - \mu.$
Since all other non-zero terms are multiples of $\alpha $
we obtain that
$$ D = -\mu (2^2 -\mu)\cdots ((2m)^2 - \mu) + O(\alpha^2).$$
By (\ref{3.32}) we have $\mu > 4m^2 + 1 .$ Therefore,
we may choose a positive number
$\varepsilon < 1/(12t) $ so that $D\neq 0 $ for each $\alpha $ such that
$|\alpha |< \varepsilon. $  
But then, the homogeneous system has only the zero solution
$A_0=0, \ldots, A_{2m} = 0. $ Since $A_0 =0 $ the $\ell^2$-sequence
$B_k = A_k, \, k=2,4,\ldots, $
satisfies (\ref{3.26}) and (\ref{3.27}),
and therefore
 $ w(x)= \sum_{k \in 2 \mathbb{N}} A_k \sin kx $
 is a non-zero odd periodic solution of (\ref{3.19}).
This means that
(\ref{3.20}) has two linearly independent periodic solutions,
which implies that (\ref{3.18}) has 
two linearly independent periodic solutions also. Thus
$\lambda $ is of multiplicity 2, so
the $n$-th zone of instability is closed, i.e.
$$ \gamma_n = \pm P_n (2m+1) \alpha^n +
 O\left (\alpha^{n+1} \right ) \equiv 0 \quad
\mbox{for} \;\; |\alpha| < \varepsilon. $$
Hence $P_n (2m+1) = 0, $  which completes the proof
of (\ref{3.15}).

Since the proof of (\ref{3.16}) is an analogue of
 the proof of (\ref{3.15}), with (\ref{3.28})--(\ref{3.31})
playing the role of (\ref{3.23})--(\ref{3.27}),
we omit it. 
\end{proof}

\section{A collateral result}

By analyzing the proof of Theorem \ref{thm2}
we obtain some interesting identities.

\begin{Theorem}
\label{thm3}
(a)  If $k$ and  $m, \; 1 \leq k \leq m, $ are fixed, then
\begin{equation}
\label{4.1}
\sum  \prod_{s=1}^k
(m^2 -i_s^2)   
= \sum_{1\leq j_1 < \cdots < j_k \leq m}
\prod_{s=1}^k (2j_s -1)^2,  
\end{equation}
where the left sum is over all indices $ i_1, \ldots, i_k $ such that
$$ -m <i_1 <\cdots <i_k <m ,
\quad |i_s - i_r | \geq 2 \;\;\text{if} \;\; s \neq r.  $$

(b)  If $k$ and  $m, \; 1 \leq k \leq m-1, $ are fixed, then
\begin{equation}
\label{4.2}
\sum \prod_{s=1}^k
\left [(2m-1)^2 -(2i_s -1)^2 \right]
= \sum_{1\leq j_1 < \cdots < j_k \leq m-1}
\prod_{s=1}^k (4j_s)^2, 
\end{equation}
where the left sum is over all indices $ i_1, \ldots, i_k $ such that
$$ -m +1 <i_1 <\cdots <i_k < m ,
\quad |i_s - i_r | \geq 2 \;\;\text{if} \;\; s \neq r.  $$
\end{Theorem}

{\em Remark.}
The terms in (\ref{4.1}) and (\ref{4.2})
look to be similar to the terms in the identity
conjectured by V. Kac and M. Wakimoto \cite{KW}
and proved by S. Milne \cite{Mil1},
and later by D. Zagier \cite{Z};
see details and further bibliography
in \cite{Mil2},
in particular, Sect. 7 and Cor. 7.6, pp. 120-121.
Our asymptotic analysis involves eigenvalues of Schr\"odinger
operators. This occurrence of eigenvalues suggests a possible link
with advanced determinant calculus
developed by G. Andrews (see C . Krattenthaler \cite{Kr}
and references there)
and Hankel determinants in S. Milne \cite{Mil2}. \vspace{2mm}

\begin{proof}
These formulae come
from the combinatorial meaning of the coefficients
of the polynomials $P_n (t) $ in the proof of Theorem \ref{thm2}.
We give details of the proof for (\ref{4.1}) only; the proof of 
(\ref{4.2}) is the same.

Let $D_{2(m-k)} $ denote the coefficient of $P_{2m} (t) $
in front of $t^{2(m-k)}.$
By (\ref{3.15}) we obtain that
\begin{equation}
\label{4.3}
D_{2(m-k)} = (-1)^k C_{2m}
 \sum_{1\leq j_1 < \cdots < j_k \leq m}
(2j_1 -1)^2 \cdots (2j_k -1)^2 
\end{equation}

On the other hand $P_{2m} (t) $ is defined by (\ref{3.11})
as a sum of $t$-monomials such that each monomial
corresponds to a walk $(j_1, \ldots, j_k)$ with positive steps
of length 2 or 4. Moreover, by (\ref{3.3}),
the degree of each such monomial
equals the number of steps of length 2.
Thus we obtain, by (\ref{3.11}), that
\begin{equation}
\label{4.5}
D_{2(m-k)} \alpha^{2m} t^{2(m-k)} =
2 \sum_{\nu \in J_s} \frac{(-2\alpha t)^{2(m-k)} (-\alpha^2)^k}
{((2m)^2 -(\nu_1)^2) \cdots ((2m)^2 -(\nu_s)^2)},
\end{equation}
where $ s= 2m-k-1$ and  $ J_s$ is the set
of all $s$-tuples of indices $  (\nu_1, \ldots, \nu_s)$ 
that correspond to walks from $-2m$ to $2m$ with
$2m-2k $  
steps of length 2 and $k$ steps of length 4.

Each term
of the sum in (\ref{4.5})
may be rewritten in the form
$$
\frac{(-1)^k 2^{2(m-k)} \alpha^{2m}t^{2(m-k)}}
{\prod_{i=1}^{2m-1} [(2m)^2 - (-2m +2i)^2]}
((2m)^2 - (2i_1)^2) \cdots ((2m)^2 - (2i_k)^2),
$$
where the $k$-tuple $(2i_1, \ldots, 2i_k)$
complements the $(2m-k-1)$-tuple
$  (\nu_1, \ldots, \nu_s)$
to the $(2m-1)$-tuple
$(-2m + 2i)_{i=1}^{2m-1}.$
Thus (\ref{4.5}) implies, in view of (\ref{3.12}), 
\begin{equation}
\label{4.7}
D_{2(m-k)} =(-1)^k  C_{2m}
\sum_{-m <i_1 <\cdots <i_k <m} (m^2 -i_1^2)\cdots (m^2 - i_k^2).
\end{equation}
Obviously, (\ref{4.3}) and (\ref{4.7})
imply (\ref{4.1}).
\end{proof}

\section{The asymptotics of spectral gaps for large $n$}

In this section we prove the following theorem.

\begin{Theorem}
\label{thm4}
Let
$\gamma_n $ be the $n$-th spectral gap of the Hill operator
\begin{equation}
\label{a0}
Ly = - y^{\prime \prime} - (4 \alpha t \cos 2x + 2 \alpha^2 \cos 4x ) y,
\end{equation}
where either both $\alpha $ and $ t\neq 0, $ are real,
or both are pure imaginary numbers.
Then the following asymptotic formulae hold for fixed
$ \alpha, t$ and $n \to \infty:$
for even $n$
\begin{equation}
\label{a1}
\gamma_n =  \frac{8 |\alpha|^n}{2^n [(n-2)!!]^2}
\left | \cos \left (\frac{\pi}{2} t \right ) \right |
\left [ 1 + O \left ( \frac{\log n }{ n }\right ) \right ],
\end{equation}
and for odd $n$
\begin{equation}
\label{a2}
\gamma_n =
 \frac{8 |\alpha|^n}{2^n [(n-2)!!]^2}
\frac{2}{\pi} \left | \sin \left ( \frac{\pi}{2} t \right )  \right |
\left [ 1 + O \left ( \frac{\log n }{ n }\right ) \right ],
\end{equation}
where
$$
(2m-1)!! = 1\cdot 3 \cdots (2m-1), \qquad (2m)!! = 2\cdot 4 \cdots (2m).
$$
\end{Theorem}

\begin{proof}
The case where $\alpha =0 $ is trivial (then $\gamma_n \equiv 0 $),
so we assume that $\alpha \neq 0.$
For convenience the proof is divided into several steps. \vspace{3mm}

{\em Step 1.} Consider all possible walks from $-n$ to $n. $
Each such walk is determined by the sequence of
its steps
\begin{equation}
\label{a3}
x = (x_1, \ldots, x_{\nu +1}),
\end{equation}
or by its vertices
\begin{equation}
\label{a4}
j_s = -n + \sum_{k=1}^s x_k, \quad s = 1, \ldots, \nu. 
\end{equation}
Of course, if we know the vertices $j_1, \ldots,  j_\nu $ then the
corresponding steps are given by
the formula
$$ x_1 = n + j_1; \;\; x_k = j_k - j_{k-1}, \; k= 2, \ldots, \nu; \;\;
x_{\nu+1} = n - j_\nu. $$

In what follows we identify each walk with the sequence of its steps.
Let $X$ denotes the set of all walks from $-n$ to $n$
that have no vertices $\pm n,$ and no zero steps.
For each $x = (x_s)_{s=1}^{\nu+1} \in X $
and each $z \in \mathbb{R} $ set
\begin{equation}
\label{a5}
B_n (x,z) = \frac{V(x_1 ) \cdots V(x_{\nu+1})}
{(n^2 - j_1^2  + z) \cdots (n^2 - j_\nu^2 + z )},
\end{equation}
where $j_s $ are given by (\ref{a4}).
With these notations part (b) of Theorem~\ref{thm1} gives
\begin{equation}
\label{a6}
\gamma_n = 2  \left |\sum_{x \in X}  B_n (x,z) \right |
\left (1 + O \left ( \frac{1}{n^2}\right )\right ),
\end{equation}
where $z= z_n$ depends on $n,$ but $ |z|<1. $

In particular, the same formula holds for the operator (\ref{a0}).
Moreover, since in that case
$$ V(m) = 0 \;\; \mbox{if} \;\; m\neq \pm 2, \pm 4, $$
it is enough to take 
into account only the walks with steps $ \pm 2 $ and $ \pm 4,$
so further we may think that $X$ denotes the set of all walks
from $-n $ to $n $ with steps $ \pm 2 $ and $ \pm 4.$ \vspace{3mm}

{\em Step 2.}  Let $X^+ $ denote the set of all walks from
$-n $ to $n$ with positive steps equal to 2 or 4.
By the proof of Theorem \ref{thm2}
(see the text from (\ref{3.8}) to (\ref{3.12}))
we know that for even $n$
\begin{equation}
\label{a8}
2\sum_{\xi \in X^+} B_n (\xi,0) =
\frac{8\alpha^n}{2^n [(n-1)!]^2} \prod_{k=1}^{n/2}
\left ( t^2 - (2k-1)^2 \right ),
\end{equation}
and for odd $n$ 
\begin{equation}
\label{a9}
2\sum_{\xi \in X^+} B_n (\xi,0) =
\frac{- 8\alpha^n t}{2^n [(n-1)!]^2}
\prod_{k=1}^{(n-1)/2}
\left ( t^2 - (2k)^2 \right ).
\end{equation}
Theorem \ref{thm2} says that the sums (\ref{a8}) and (\ref{a9})
give the main part of the asymptotics of $ \gamma_n $
as $\alpha \to 0. $ 
We are going to prove, for fixed $\alpha $ and $t,$ that the
same expressions give the asymptotics of
$ \gamma_n $ for large $n.$ 

Since for even $n$ we have
$$
\cos \left ( \frac{\pi}{2} t \right ) =
\prod_{k=1}^\infty \left ( 1- \frac{t^2}{(2k-1)^2} \right ) =
\prod_{k=1}^{n/2} \left ( 1- \frac{t^2}{(2k-1)^2} \right )
\left [1 + O\left (\frac{1}{n} \right) \right ],
$$
and for odd $n$
$$
\sin \left ( \frac{\pi}{2}t \right ) = \frac{\pi t}{2}
\prod_{k=1}^\infty \left ( 1- \frac{t^2}{(2k)^2} \right ) =\frac{\pi t}{2}
\prod_{k=1}^{(n-1)/2} \left ( 1- \frac{t^2}{(2k)^2} \right )
\left [1 + O\left (\frac{1}{n} \right) \right ],
$$
(\ref{a8})  can be rewritten for even $n$ as
\begin{equation}
\label{a11}
2\sum_{\xi \in X^+} B_n (\xi,0) =
 \frac{\pm 8 \alpha^n}{2^n [(n-2)!!]^2}
\cos \left (\frac{\pi}{2} t \right )
\left [1 + O\left (\frac{1}{n} \right) \right ],
\end{equation}
while (\ref{a9}) gives for odd $n$ 
\begin{equation}
\label{a12}
2\sum_{\xi \in X^+} B_n (\xi,0) =
 \frac{\pm 8 \alpha^n}{2^n [(n-2)!!]^2} \cdot
\frac{2}{\pi} \sin \left ( \frac{\pi}{2} t \right )
\left [1 + O\left (\frac{1}{n} \right) \right ].
\end{equation}

In view of (\ref{a6}), (\ref{a11}) and (\ref{a12}),
to accomplish the proof of Theorem~\ref{thm4} we need to show that
$$ \sum_{x \in X} B_n (x,z)  \asymp  \sum_{\xi \in X^+} B_n (\xi,0)
\qquad \text{as} \quad n \to \infty,
$$
where $ z= z_n $ with $ |z| < 1. $
In Lemma \ref{lema2} below it is proven that
$$
 \sum_{\xi \in X^+} B_n (\xi,z) \asymp \sum_{\xi \in X^+} B_n (\xi,0).
 $$
 The remaining part of the proof shows that
 $ \sum_{x \in X\setminus X^+} B_n (x,z) $ is relatively small
in comparison with
$ \sum_{\xi \in X^+} B_n (\xi,z) . $ \vspace{3mm}

{\em Step 3.} Two technical lemmas.

\begin{Lemma}
\label{lema1}
\begin{equation}
\label{a13}
\frac{\log n}{2n} \leq 
\sum_{0< i< n} \frac{1}{n^2 - (n-2i)^2} \leq \frac{1+\log n}{2n}.
\end{equation}
\begin{equation}
\label{a14}
\sum_{i\neq 0,n} \left | \frac{1}{n^2 - (n-2i)^2} \right |
\leq \frac{1+\log n}{n}.
\end{equation}
\end{Lemma}
The proof is elementary, and it is omitted.

\begin{Lemma}
\label{lema2}
If $\xi \in X^+ $
and $n \geq 3 $ then for $z \in [0,1) $
\begin{equation}
\label{a15}
1-z \frac{\log n}{n} \leq \frac{B_n (\xi,z)}{B_n (\xi,0)}
\leq  1- z \frac{\log n}{4n}, 
\end{equation}
and for $z \in (-1,0]$
\begin{equation}
\label{a16}
1+|z| \frac{\log n}{2n} \leq \frac{B_n (\xi,z)}{B_n (\xi,0)}
\leq 1+ |z|  \frac{2\log n}{n} 
\end{equation}
\end{Lemma}

\begin{proof}
By (\ref{a5}),
$$ \frac{B_n (\xi,z)}{B_n (\xi,0)} =
\prod_{0<i<n} \left (1+ \frac{z}{n^2 - (n^2 - 2i)^2} \right )^{-1}.
$$
One can easily see (since $1+x \leq e^x ,\; \forall x \in \mathbb{R}$),
that
if either
$y_i \geq 0, \, i =1, \ldots, m, $
or $ y_i \in [-1,0], \, i =1, \ldots, m, $
 then
\begin{equation}
\label{a17}
1 + \sum_{i=1}^{m} y_i \leq (1+y_1 ) \cdots (1+y_{m})
\leq \exp \left ( \sum_{i=1}^{m} y_i \right ).
\end{equation}
Now from (\ref{a17}) it follows that
if $y_i \in (-1, 0] \; \forall i, $
or $y_i \geq 0 \; \forall i, $  then
\begin{equation}
\label{a18}
1 - \sum_{i=1}^{m} y_i \leq
\exp \left ( - \sum_{i=1}^{m} y_i \right ) \leq
\left ( \prod_{i=1}^{m} (1+y_i ) \right )^{-1}
\leq \left ( 1+ \sum_{i=1}^{m} y_i \right )^{-1}.
\end{equation}

We use the inequalities (\ref{a18}) with
$y_i = z/(n^2 - (n-2i)^2 ), \; i=1, \ldots, n-1. $
If $z \in [0,1),$ then by (\ref{a18}) and (\ref{a13}),
we have
$$
1- z\frac{1+ \log n}{2n} \leq \frac{B_n (\xi,z)}{B_n (\xi,0)}
\leq \left (1 + z \frac{\log n}{2n} \right )^{-1}.
$$
Taking into account that
$(1+\varepsilon )^{-1} < 1- \varepsilon /2 $
with
$\varepsilon = z\log n /(2n) <1, $
we obtain (\ref{a15}).

If $z \in (-1,0]$ then, again by (\ref{a18}) and (\ref{a13}),
we obtain
$$
1 + |z| \frac{\log n}{2n} \leq \frac{B_n (\xi,z)}{B_n (\xi,0)}
\leq \left (1 - |z| \frac{1+\log n}{2n} \right )^{-1}.
$$
If $ n\geq 3 $ then, with $\varepsilon = |z| (1+ \log n )/(2n) \leq 1/2, $
we have
$$
(1-\varepsilon )^{-1} \leq 1 + 2\varepsilon \leq 
1+ 2 |z| \frac{1+\log n}{2n} \leq 1 + |z| \frac{2\log n}{n},
$$
which proves (\ref{a16}).
\end{proof}

{\em Step 4.}
This step contains some constructions and inequalities 
that are crucial for the estimate of
the sum
$\sum_{x\in X\setminus X^+} B_n (x,z). $
For each $ \xi \in X^+ $ let
$X_{\xi} $ denote the set of all walks $ x \in X\setminus X^+ $
such that each vertex of $\xi $ is a vertex of $x$ also.
It is easy to see that
\begin{equation}
\label{k1}
 X \setminus X^+ = \bigcup_{\xi \in X^+} X_{\xi},
\end{equation}
i.e., for each $x\in X\setminus X^+ $
there exists a $\xi \in X^+ $
such that each vertex of $\xi $ is a vertex of $x.$
Indeed, fix a walk $x \in X\setminus X^+. $  If
$(v_k)_{k=1}^r $ is the sequence of its vertices, then we
define a strictly increasing subsequence $(v_{k_s})_{s=1}^\nu $
of it as follows:
\begin{equation}
\label{k2}
k_1 = \min \{k: \; -n < v_k < n\}, \quad
k_s = \min \{k: \; v_{k_{s-1}} < v_k < n\}. 
\end{equation}
For some $\nu \leq r $
one would get that such a choice cannot continue anymore
and stop:
this gives the last term $ v_{k_\nu} $
of the subsequence.
Since each step of $x$ is equal to $\pm 4$ or $\pm 2, $
the distance between every two
consecutive terms
of the subsequence  $(v_{k_s})_{s=1}^\nu $
is equal to 2 or 4, and the same is true for the
differences $ v_{k_1} - (-n) $ and $n- v_{\nu}. $
Thus $(v_{k_s})_{s=1}^\nu $ is the sequence of the vertices of
a walk $ \xi \in X^+$
such that each vertex of $\xi $ is a vertex of $x.$

For each $\xi \in X^+ $ and 
$ \mu \in \mathbb{N} $ let
$X_{\xi,\mu}  $ be the set of all 
$x \in  X_\xi $
such that $x$ has $\mu $ more vertices than $\xi.$
Then we have
\begin{equation}
\label{k3}
X_\xi = \bigcup_{\mu =1}^\infty  X_{\xi,\mu}.
\end{equation}

Moreover, for each $\mu $-tuple $(i_1, \ldots, i_\mu ) $ of
integers in
$$
I_n = \left ( n + 2 \mathbb{Z} \right ) \setminus \{ \pm n \}
$$
we define
$X_{\xi} (i_1, \ldots, i_\mu) $ as the set of all walks $x$
with $\nu + 1 + \mu $ steps
such that $ (i_1, \ldots, i_\mu) $ and
the sequence of the vertices of $\xi$ are complementary subsequences of
the sequence of the vertices of $x.$
Then
\begin{equation}
\label{k4}  
X_{\xi,\mu} = \bigcup_{(i_1, \ldots i_\mu ) \in (I_n)^\mu }
X_{\xi} (i_1, \ldots, i_\mu).
\end{equation}

\begin{Lemma}
\label{lema4}
Under the above notations,
for each walk $ \xi \in X^+ $ and
each $\mu$-tuple $ (i_1, \ldots i_\mu ) \in (I_n)^\mu, $
\begin{equation}
\label{k5}
\# \, X_{\xi} (i_1, \ldots, i_\mu) \leq 5^\mu.
\end{equation}
\end{Lemma}

\begin{proof}
Fix $\xi \in X^+ $ and let $(j_s)_{s=1}^\infty $ be
the sequence of its vertices.
If $ x \in X_{\xi} (i_1, \ldots, i_\mu) $
then the sequence of the vertices of $x$
may be obtained by adding
$i_1, \ldots, i_\mu, $ one by one,
at appropriate places, as new terms to
the sequence $ (j_s)_1^\nu. $ 

For convenience we put
$j_0 = -n $ and $ j_{\nu +1 } = n. $
The integer $i_1 $ could appear as a vertex in a prospective walk
$ x \in X_{\xi} (i_1, \ldots, i_\mu) $
after some $\xi$-vertex $j_{s_1}. $
But then
$ i_1 - j_s = \pm 2, \pm 4,$
so there are only four possible choices for $s_1, $ i.e.,
at most 4 choices where to place $i_1. $

If $ i_1, \ldots, i_m ,\, 1 \leq m < \mu $ 
have been properly placed, then the next vertex
$i_{m+1} $ could appear immediately after $i_m $ (one option)
or after some $\xi$-vertex $j_{s_{m+1}}, $
but then 
$ i_{m+1} - j_{s_{m+1}} = \pm 2, \pm 4,$
so there are at most 5 options
for the spot where $i_{m+1} $ could be placed.
Thus, for each $k, \, 1\leq k \leq \mu ,$ 
there are at most $5$ choices for a place for $i_k, $
and therefore,
the cardinality of
$X_{\xi} (i_1, \ldots, i_\mu) $ does not exceed  $ 5^\mu.$

\end{proof}

{\em Remark.} We could use instead of
$X_\xi (i_1, \ldots, i_\mu) $ its subset
$X^\prime_\xi (i_1, \ldots, i_\mu) $ of the walks $x$
which lead to $\xi $ after restructuring defined by (\ref{k2}).
Then one can show that 
$\# \, X^\prime_{\xi} (i_1, \ldots, i_\mu) \leq 3^\mu.$ \vspace{2mm}

\begin{Lemma}
\label{lema5} 
If
$\xi \in X^+ $ and $ |z| \leq 1 $
then there exists $n_1 $ such that  for $ n \geq n_1 $
\begin{equation}
\label{k7}
\sum_{x \in X_\xi}  |B_n (x,z)|
\leq  |B_n (\xi,z)|
\cdot \frac{K \log n}{n} 
\end{equation}
where $ K = 160 (|t| + |\alpha |)^2.$
\end{Lemma}

\begin{proof}
By (\ref{k3}) 
\begin{equation}
\label{k8}
\sum_{x \in X_\xi}  |B_n (x,z)|
= \sum_{\mu =1}^\infty
\sum_{x \in X_{\xi,\mu}}  |B_n (x,z)|,
\end{equation}
We are going to show that
\begin{equation}
\label{k10}
\sum_{x \in X_{\xi,\mu}}  |B_n (x,z)|
\leq  |B_n (\xi,z)|
\left (  \frac{20 C \log n}{n} \right )^\mu
\end{equation}
where $ C = 4 (|t| + |\alpha |)^2.$
If (\ref{k10}) is proven, then with
$n_1 $ chosen so that $(20 C \log n) /n \leq 1/2 $
for $n \geq n_1 $ one would obtain, by (\ref{k8}),
$$
\sum_{x \in X_\xi} \frac{|B_n (x,z)|}{|B_n (\xi,z)|}
 \leq \sum_{\mu =1}^\infty
\left (  \frac{20 C \log n}{n} \right )^\mu
\leq  \frac{40 C \log n}{n}, 
$$
i.e., (\ref{k7}) would hold with $K = 40 C.$

By (\ref{k3}),
\begin{equation}
\label{k11}
\sum_{X_{\xi, \mu}} |B_n (x,z)| \leq
\sum_{(i_1, \ldots, i_\mu)}
\sum_{X_\xi (i_1, \ldots, i_\mu)} |B_n (x,z)|,
\end{equation}
where the first sum on the right is taken over all 
 $\mu $-tuples $(i_1, \ldots, i_\mu ) $ of
integers $ i_s \in n + 2\mathbb{Z} $  
such that $i_s \neq \pm n. $

Fix $(i_1, \ldots, i_\mu ). $
If $ x \in X_\xi(i_1, \ldots, i_\mu), $
then, in view of (\ref{a5}),
\begin{equation}
\label{k12}
\frac{B_n (x,z)}{B_n (\xi,z)} = \frac{\prod_k V(x_k )}{\prod_s V(\xi_s)}
\cdot \frac{1}{(n^2 - i_1^2 +z) \cdots (n^2 - i_\mu^2 +z)}.
\end{equation}
If each step of $\xi $ is a step of $x,$ then
(since $V(x_k ) = - 2 \alpha t $ if $ x_k = \pm 2, $ and
$V(x_k ) = -  \alpha^2 $ if $ x_k = \pm 4 $)
\begin{equation}
\label{k13}
\frac{\prod_k |V(x_k )|}{\prod_s | V(\xi_s)|}  \leq C^\mu
\end{equation}
(because $x$ has $\mu $ steps more).
The same is true in the general case also.
Indeed, let $(j_s)_{s=1}^\nu $ are
the vertices of $\xi,$ and let us put for convenience
$ j_0 = -n $ and $ j_{\nu+1} = n.$
Since each vertex of $\xi $ is a vertex of $x,$
for each $s, $  $ 1 \leq s \leq \nu + 1, $
$$\xi_s = j_s - j_{s-1} = \sum_{k \in J_s} x_k
$$
where $x_k, \, k \in J_s ,$ are the steps of $x$ between
the vertices
$j_{s-1} $ and $j_s.$
Fix an $s, \, 1 \leq s \leq \nu +1. $
If $ \xi_s =2, $ then
there is a step $x_{k^*}, \, k^* \in J_s, $
such that $|x_{k^*} | =2 $
(otherwise $\xi_s $ would be a multiple of 4).
Therefore, $|V(\xi_s)| = |V(x_{k^*})|,$
and
\begin{equation}
\label{k15}
\frac{\prod_{J_s} |V(x_k)|}{|V(\xi_s)|}  \leq
C^{b_s -1}, \quad \text{where} \;\; b_s := \# J_s.
\end{equation}
Suppose $\xi_s = 4.$ If there is $k_* \in J_s $ with
$ |x_{k_*}|= 4,$
then  $|V(\xi_s)| = |V(x_{k_*})|,$
so (\ref{k15}) holds.
Otherwise, there are $k^\prime, k^{\prime \prime} \in J_s $
such that $ |x_{k^\prime}| =|x_{k^{\prime \prime}}|= 2, $
and therefore,
$$
\frac{|V(x_{k^\prime}) V(x_{k^{\prime \prime}})|}
{|V (\xi_s)|} = \frac{4|\alpha|^2 |t|^2}{|\alpha|^2}
= 4 |t|^2 \leq C,
$$
which implies (\ref{k15}).

Since
$ \sum_s (b_s -1) = \mu, $
(\ref{k15}) yields (\ref{k13}).
Therefore, using the elementary inequality
$$
|n^2 -i^2 +z|^{-1} \leq 2 |n^2 -i^2|^{-1}, \quad i \neq \pm n,
\;\; |z| \leq 1, 
$$
we obtain, by (\ref{k12}) and (\ref{k13}), that
$$
\frac{|B_n (x,z)|}{|B_n (\xi,z)|} \leq \frac{(2C)^\mu}
{|n^2 - i_1^2| \cdots |n^2 -i_\mu^2|},
\quad x \in X_\xi (i^1, \ldots, i_\mu).
$$
Now, by Lemma \ref{lema4},
$$
\sum_{x \in X_\xi (i^1, \ldots, i_\mu)}
\frac{|B_n (x,z)|}{|B_n (\xi,z)|} \leq
\frac{(10C)^\mu}{|n^2 - i_1^2| \cdots |n^2 -i_\mu^2|}.
$$
Thus, by (\ref{k11}) and Lemma \ref{lema2},
$$
\sum_{X_{\xi, \mu}} \frac{|B_n (x,z)|}{|B_n (\xi,z)|} \leq
\sum_{(i_1, \ldots, i_\mu)}
\frac{(10C)^\mu}{|n^2 - i_1^2| \cdots |n^2 -i_\mu^2|} 
\leq   \left ( \sum_{i \in (n+ 2\mathbb{Z})\setminus \{\pm n \}}
\frac{10C}{|n^2 - i^2|} \right )^\mu     $$
$$\leq
(10C)^\mu \left ( \frac{1 + \log n}{n} \right )^\mu \leq
\left ( \frac{(20C) \log n}{n} \right )^\mu,
$$
i.e., (\ref{k10}) holds.
This completes the proof of Lemma \ref{lema5}.

\end{proof}

{\em Step 5.}
This step completes the proof of Theorem \ref{thm4}
for even $n.$

If $ t = 2k-1, \; k=1,2, \ldots, $ then,
by Theorem 11 in \cite{DM12}, 
$\gamma_n = 0 $ for $ n> 2k,$
thus (\ref{a1}) holds.

Suppose that $\alpha $ and $t$ are nonzero real numbers,
and $t \neq 2k-1, \, k \in \mathbb{N}.$

By (\ref{k1}), $X\setminus X^+ = \bigcup_{\xi \in X^+} X_{\xi}.$
Let us choose disjoint sets $X^\prime_\xi \subset X_\xi $ so that
\begin{equation}
\label{k19}
X^\prime = \bigcup_{\xi \in X^+} X^\prime_\xi.
\end{equation}
Then
\begin{equation}
\label{m70}
\sum_{x \in X \setminus X^+} B_n (x,z) =
\sum_{\xi \in X^+} \left ( \sum_{x\in X^\prime_\xi} B_n (x,z),
\right )
\end{equation}
and therefore,
by (\ref{m70}),
we have
\begin{equation}
\label{m73}
\sum_{x\in X} B_n (x,z)
=\sum_{\xi \in X^+} 
\left (
B_n (\xi,z) +
\sum_{x \in X^\prime_\xi} B_n (x,z)
\right )  = \Sigma^- + \Sigma^+,
\end{equation}
where
\begin{equation}
\label{m75}
\Sigma^-  = \sum_{\xi: B_n (\xi, 0)<0} \cdots, \qquad
\Sigma^+  = \sum_{\xi: B_n (\xi, 0)>0} \cdots.
\end{equation}
Set
\begin{equation}
\label{m76}
\sigma_n^-  = \sum_{\xi: B_n (\xi, 0)<0} B_n (\xi,0), 
\qquad
\sigma_n^+  = \sum_{\xi: B_n (\xi, 0)>0} B_n (\xi,0), 
\end{equation}
and
\begin{equation}
\label{m76a}
\sigma_n = \sum_{\xi \in X^+} B_n (\xi,0) =
\sigma_n^- + \sigma_n^+ .
\end{equation}

By 
(\ref{a15}) and (\ref{a16}) 
in Lemma \ref{lema2},
and by (\ref{k7}), 
we obtain,
for each $\xi $ with $B_n (\xi,0) <0, $
$$
\left [ 1+ C \frac{\log n}{n} \right ] B_n (\xi,0) \leq
B_n (\xi,z) +
\sum_{x \in X^\prime_\xi} B_n (x,z)
\leq
\left [ 1- C \frac{\log n}{n} \right ] B_n (\xi,0),
$$
and therefore,
\begin{equation}
\label{m77}
\left [ 1+ C \frac{\log n}{n} \right ] \sigma_n^- \leq
\Sigma^- \leq
\left [ 1- C \frac{\log n}{n} \right ] \sigma_n^-,
\end{equation}
where the constant $C> 0 $ depends on $\alpha $ and $t$ only.

In an analogous way it follows 
\begin{equation}
\label{m78}
\left [ 1- C \frac{\log n}{n} \right ] \sigma_n^+ \leq
\Sigma^+ \leq
\left [ 1+ C \frac{\log n}{n} \right ] \sigma_n^+
\end{equation}
(with the same constant $C,$ otherwise we may take a greater constant $C$
in (\ref{m77})).

Now (\ref{m73}) - (\ref{m78}) yield
$$
-C(|\sigma^-_n| + \sigma^+_n) \frac{\log n}{n} \leq
 \sum_{x \in X} B_n (x,z) - \sigma_n
\leq C(|\sigma^-_n| + \sigma^+_n) \frac{\log n}{n},
$$ 
and therefore,
\begin{equation}
\label{m79}
\left | \frac{1}{\sigma_n} \sum_{x \in X} B_n (x,z) -1
\right | \leq C \, \frac{|\sigma^-_n| + \sigma^+_n}{|\sigma_n|}
 \cdot \frac{\log n}{n}.
\end{equation}

In view of (\ref{a8}) and (\ref{m76a}),
\begin{equation}
\label{m81}
\sigma_n =
\sum_{\xi \in X^+} B_n (\xi,0) =
\frac{ 4\alpha^n}{2^n [(n-1)!]^2} \prod_{k=1}^{n/2}
\left ( t^2 - (2k-1)^2 \right ),
\end{equation}
thus $\sigma_n \neq 0$ (since $t \neq 2k-1, \, k \in \mathbb{N} $).
By (\ref{m76}),
\begin{equation}
\label{m82}
|\sigma_n^-|+ \sigma_n^+ 
=\frac{ 4 \alpha^n}{2^n [(n-1)!]^2} \prod_{k=1}^{n/2}
\left ( t^2 + (2k-1)^2 \right ),
\end{equation}
so (for $n > |t| $)
\begin{eqnarray}
\nonumber
\frac{|\sigma^-_n| + \sigma^+_n}{|\sigma_n|}
= \frac{\prod_{k=1}^{n/2} \left ( t^2 + (2k-1)^2 \right ) }
{ \left |\prod_{k=1}^{n/2} \left ( t^2 - (2k-1)^2 \right ) \right | } =
\frac{\prod_{k=1}^{n/2} \left ( 1+ \frac{t^2}{(2k-1)^2} \right ) }
{\prod_{k=1}^{n/2} \left | 1- \frac{t^2}{(2k-1)^2} \right | } \\
{} \label{m84} \\
\nonumber
\leq
\frac{\prod_{k=1}^\infty \left ( 1+ \frac{t^2}{(2k-1)^2} \right ) }
{\prod_{k=1}^\infty \left | 1- \frac{t^2}{(2k-1)^2} \right | }
=\left | \frac{\cosh \left (\frac{\pi}{2}t \right )}
{\cos \left (\frac{\pi}{2}t \right )} \right |.
\end{eqnarray}
Hence, by (\ref{m79}),
\begin{equation}
\label{m85}
\sum_{x \in X} B_n (x,z)
= \sigma_n \left [ 1 + O\left ( \frac{\log n}{n} \right ) \right ].
\end{equation}

If $\alpha $ and $t$ are pure imaginary,
then the situation is more simple because
$B(\xi, 0) > 0 $ for each $\xi \in X^+. $
Thus we have
$ \sigma_n = \sigma^+_n >0 $
and $ \sum_{x \in X} = \Sigma^+,$
so (\ref{m78}) yields immediately
(\ref{m85}).

Finally, by (\ref{a8}) and (\ref{a11}),
(\ref{m85}) implies
\begin{equation}
\label{m86}
2\sum_{x \in X} B_n (x,z)
=
 \frac{\pm 8\alpha^n}{2^n [(n-2)!!]^2}
\cos \left (\frac{\pi}{2} t \right ) 
\left [ 1 + O\left ( \frac{\log n}{n} \right ) \right ].
\end{equation}
In view of (\ref{a6}) the estimate (\ref{m86})
proves Theorem \ref{thm4}
for even $n,$ 
i.e., (\ref{a1}) holds. \vspace{3mm}

{\em Step 6.}
This step completes the proof of Theorem \ref{thm4}
for odd $n.$ 

If $ t = 2k, \; k=0, 1,2, \ldots, $ then,
by Theorem 11 in \cite{DM12}, 
$\gamma_n = 0 $ for $ n> 2k+1,$
thus (\ref{a2}) holds.  

Suppose that $\alpha $ and $t$ are non-zero real numbers,
and $t\neq 2k, \, k \in \mathbb{N}. $
Using the same argument and notations as in Step 5, we obtain
(see (\ref{k19})-(\ref{m79}))
that (\ref{m79}) holds for odd $n.$

In view of (\ref{a9}) and (\ref{m76a}),
\begin{equation}
\label{am81}
\sigma_n =
\frac{- 4\alpha^n t}{2^n [(n-1)!]^2}
\prod_{k=1}^{(n-1)/2}
\left ( t^2 - (2k)^2 \right ),
\end{equation}
thus $\sigma_n \neq 0$ (since $t \neq 2k, \, k \in \mathbb{N} $).
By (\ref{m76}),
$$
|\sigma_n^-|+ \sigma_n^+
=\frac{ 4 \alpha^n |t|}{2^n [(n-1)!]^2} \prod_{k=1}^{(n-1)/2}
\left ( t^2 + (2k)^2 \right ),
$$
so (for $n > |t| $)
\begin{eqnarray}
\nonumber
\frac{|\sigma^-_n| + \sigma^+_n}{|\sigma_n|}
= \frac{\prod_{k=1}^{(n-1)/2} \left ( t^2 + (2k)^2 \right ) }
{ \left |\prod_{k=1}^{(n-1)/2} \left ( t^2 - (2k)^2 \right ) \right | } =
\frac{\prod_{k=1}^{(n-1)/2} \left ( 1+ \frac{t^2}{(2k)^2} \right ) }
{\prod_{k=1}^{(n-1)/2} \left | 1- \frac{t^2}{(2k)^2} \right | }\\
{} \label{om84}\\
\leq
\frac{\prod_{k=1}^\infty \left ( 1+ \frac{t^2}{(2k)^2} \right ) }
{\prod_{k=1}^\infty \left | 1- \frac{t^2}{(2k)^2} \right | }
= \left | \frac{\sinh \left (\frac{\pi}{2}t \right )}
{\sin \left (\frac{\pi}{2}t \right )} \right |. \nonumber
\end{eqnarray}
Hence, by (\ref{m79}),
\begin{equation}
\label{am85}
\sum_{x \in X} B_n (x,z)
= \sigma_n \left [ 1 + O\left ( \frac{\log n}{n} \right ) \right ].
\end{equation}

If $\alpha $ and $t$ are pure imaginary,
then
$B(\xi, 0) > 0 $ for each $\xi \in X^+. $
Thus we have
$ \sigma_n = \sigma^+_n >0 $
and $ \sum_{x \in X} = \Sigma^+,$
so (\ref{m78}) yields immediately
(\ref{am85}).

Finally, by (\ref{a9}) and (\ref{a12}),
(\ref{am85}) implies
\begin{equation}
\label{am86}
2\sum_{x \in X} B_n (x,z)
=
 \frac{\pm 8\alpha^n t}{2^n [(n-2)!!]^2}
\frac{2}{\pi}\sin \left (\frac{\pi}{2} t \right ) 
\left [ 1 + O\left ( \frac{\log n}{n} \right ) \right ].
\end{equation}
In view of (\ref{a6}) the estimate (\ref{am86})
proves Theorem \ref{thm4}
for odd $n,$ 
i.e., (\ref{a2}) holds.

\end{proof}

\section{Comments and generalizations}

1. In Sections 3-5 we consider only two term potentials of the form
$v(x) = a \cos 2x + b \cos 4x. $
Now we would like to make some comments about the more general case
where $v$ is a real-valued trigonometric polynomial of the form
\begin{equation}	
\label{f1}
v(x) = \sum_{k=1}^K (a_k e^{2ikx} + \overline{a}_k e^{-2ikx}).
\end{equation}
For each $n \in \mathbb{N} $
let $X_n $ and $X_n^+ $ denote, respectively,
the set of all walks from $-n$ to $n$ with steps $\pm 2, \ldots, \pm 2K,$
and the subset of $X_n $ of all walks with positive steps.
With these notations the general asymptotic formula
(\ref{d21}) from Theorem \ref{thm1} becomes
\begin{equation}
\label{f2}
\gamma_n = 2  \left | \sum_{x \in X_n} B_n (x,z) \right |
\left ( 1+ O\left ( \frac{\|v\|^2}{n^2} \right ) \right ),
\quad z = z_n,
\end{equation}
where either $|z_n| \leq 4 \|v\| $ for all $n$ (if $\|v\| <1/9$),
or $ |z_n | <1 $ for large enough $n.$

Consider the parametrization
\begin{equation}
\label{f3}
a_1 = t_1 \alpha, \; a_2 = t_2 \alpha^2, \ldots
a_{K-1} = t_{K-1} \alpha^{K-1}, \; a_K = \alpha^K.
\end{equation}
Then, by part (a) of Proposition \ref{prop1},
$|z| \leq 4 \|v|| = O(|\alpha|) $ for small $\alpha,$
so (\ref{f2}) yields (as in the proof of Theorem \ref{thm2})
that
\begin{equation}
\label{f4}
\gamma_n = 2 \left | \sum_{\xi \in X^+_n} B_n (\xi,0) \right |
+ O \left (|\alpha |^{n+1} \right ).
\end{equation}

If $K=1 $ then
there is only one walk
with positive steps from $-n $ to $n,$
namely $\xi^* = (2, \ldots, 2) $
(i.e., $X_n^+ = \{\xi^* \}).$
Since
$$
B(\xi^*,0) = \frac{\alpha^n}{\prod_{j=1}^{n-1}
[n^2 - (-n +2j)^2 ]} = \frac{\alpha^n}{4^{n-1} [(n-1)!]^2},
$$
we obtain
\begin{equation}
\label{f5}
\gamma_n = 2
\left | B_n (\xi^*,0) \right | + O \left (|\alpha |^{n+1} \right ) =
\frac{|\alpha|^n}{4^{n-1} [(n-1)!]^2} (1+O (|\alpha |)),
\end{equation}
which gives the Levy-Keller's formula (\ref{i08})
for the Mathieu potential. \vspace{3mm}

2. If $K>1 $ then the computation of
$\sum_{\xi \in X_n^+} B_n (\xi, 0) $ is not trivial (unless all $t$'s vanish
and there is only one walk $\xi$
with positive steps that gives a non-zero term $B_n (\xi, 0) $).

As in the proof of Theorem \ref{thm2} one can easily see,
for each $K, $ that
\begin{equation}
\label{f6}
\sum_{x \in X_n^+} B_n (x, 0) = \alpha^n \cdot P_n (t_1, \ldots, t_{K-1}), 
\end{equation}
where $P$ is a polynomial in $ t_1, \ldots, t_{K-1}.$

Our main achievement in Theorem \ref{thm2} (where $K=2$)
is the explicit form of
the corresponding polynomials $P_n, \, n \in N.$
In Theorem \ref{thm2} we consider potentials
of the form $ v(x) = - 4 \alpha t \cos 2x - 2\alpha^2 \cos 4x, $
where $ \alpha $ and $t$ are simultaneously real or pure
imaginary,
while the potentials that comes from (\ref{f1}) for $K=2 $
are more general, namely
\begin{equation}
\label{f7}
v(x) = a_1 e^{2ix} + \overline{a}_1 e^{-2ix} +
a_2 e^{4ix} + \overline{a}_2 e^{-4ix},  \quad
a_1, a_2 \in \mathbb{C}, \; a_2 \neq 0, 
\end{equation}
or equivalently,
\begin{equation}
\label{f8}
v(x) = A \cos 2x + B \sin 2x + C \cos 4x + D \sin 4x, 
\end{equation}
where
$$ A= 2 Re \, a_1, \;\; B= -2 Im \,a_1, \quad 
   C= 2 Re \, a_2, \; \;  D= -2 Im \,a_2,
$$

Using the same parametrization as in Theorem \ref{thm2}
(it is slightly different from (\ref{f3})),
we may write each potential $v \in (\ref{f7}), (\ref{f8}) $
as
\begin{equation}
\label{f9}
v(x) = - 2 \alpha t e^{2ix} - 2 \overline{\alpha t} e^{-2ix}
-  \alpha^2 e^{4ix} -  \overline{\alpha}^2 e^{-4ix},
\end{equation}
where $ \alpha $ and $t$ are complex numbers such that
$ \alpha^2 = - a_2, \; 2 \alpha t = - a_1. $

Observe that formally the expression
$\sum_{x \in X^+_n} B_n (x, 0) $ is exactly the same that
has been used in the proof of Theorem \ref{thm2}, because
$ \overline{\alpha} $ and $\overline{t} $ would appear in
$B_n (x,0) $ only if $x$ has negative steps.
Therefore, the same argument that proves Theorem \ref{thm2}
shows that the following more general statement holds.

\begin{Theorem}
\label{thm2a}
Let $\gamma_n, \, n \in \mathbb{N} $
be the lengths of instability zones of
the Hill operator
which potential $v$ is given by (\ref{f9}), with
$\alpha, t \in \mathbb{C} $ and $ \alpha \neq 0. $
If $t$ is fixed and $ \alpha \to 0, $ then for even $n$ 
\begin{equation}
\label{f10}
\gamma_n = \left | \frac{8\alpha^n}{2^n [(n-1)!]^2} \prod_{k=1}^{n/2}
\left ( t^2 - (2k-1)^2 \right ) \right |
\left ( 1 + O(\alpha) \right ),
\end{equation}
and for odd $n$
\begin{equation}
\label{f11}
\gamma_n = \left | \frac{ 8\alpha^n t}{2^n [(n-1)!]^2}
\prod_{k=1}^{(n-1)/2}
\left ( t^2 - (2k)^2 \right ) \right | \left ( 1 + O(\alpha ) \right ).
\end{equation}
\end{Theorem}

If $K >2 $ we don't know any explicit formula for
the asymptotics of 
$\gamma_n $ as $\alpha \to 0 $ besides
the simple extensions of Theorem \ref{thm2a} and
(\ref{f5})
that one can obtain
by using the following elementary statement
(see \cite{DM9}, Prop. 20 and 24).

\begin{Proposition}
\label{prop2}
Suppose $m \in \mathbb{N}, \, m>1, $ is fixed.
Let $\gamma_n, \, n \in \mathbb{N} $
be the lengths of instability zones of
the Hill operator
with a potential $v, $
and  let $ \tilde{\gamma}_n $
be the lengths of instability zones of
the Hill operator
which potential is $\tilde{v} (x) = m^2 v(mx). $
Then  
\begin{equation}
\label{13}
\tilde{\gamma}_{mn} = m^2 \gamma_n, \qquad
\tilde{\gamma}_k = 0 \;\; \text{if} \;\; k \not \in m \mathbb{N}.
\end{equation}
\end{Proposition}

3. If we fix $\alpha $ and $t,$ then 
the proof of Theorem \ref{thm4}
(with only a slight change in its Steps 5 and 6)
proves the following more general claim.

\begin{Theorem}
\label{thm4a}
Let $\gamma_n, \, n \in \mathbb{N} $
be the lengths of instability zones of
the Hill operator
which potential $v$ is given by (\ref{f9}), with
$\alpha, t \in \mathbb{C}, \alpha \neq 0. $
Then, for fixed $ \alpha $ and $ t,$
the asymptotic formula (\ref{a1}) holds for even $n \to \infty, $
while
the asymptotic formula (\ref{a2}) holds for odd $n \to \infty. $
\end{Theorem}

But here we would like to reformulate Theorem \ref{thm4a} so that
to have the asymptotics of $ \gamma_n $ given explicitly in
terms of the coefficients $ a_1 $ and $a_2. $
In fact, we proved (see Step 2 in the proof of Theorem \ref{thm4}),
that (\ref{a8}) and (\ref{a9}) give the asymptotics of
$\gamma_n, $  respectively, for even and odd $n.$

Replacing, respectively in (\ref{a1}) and (\ref{a2}),
$\cos (\frac{\pi}{2} t) $ and $ \frac{2}{\pi} \sin (\frac{\pi}{2} t)$
with the right-hand sides of (\ref{a8}) and (\ref{a9}),
and taking into account that
$$
a_1 = - 2 \alpha t \quad  \text{and} \quad  a_2 = - \alpha^2,
$$
we obtain, for $a_2 \neq 0, $ the following theorem.

\begin{Theorem}
\label{thm4b}
Let $\gamma_n, \, n \in \mathbb{N} $
be the lengths of instability zones of
the Hill operator $Ly = - y^{\prime \prime} + v(x) y, $
where
\begin{equation}
\label{f17}
v(x) = a_1 e^{2ix} + \overline{a}_1 e^{-2ix} +
a_2 e^{4ix} + \overline{a}_2 e^{-4ix},
\end{equation}
where $ a_1, \, a_2 \in \mathbb{C}. $
Then, for even $n,$
\begin{equation}
\label{f18}
\gamma_n =
\left | \frac{8}{2^n [(n-1)!]^2} \prod_{k=1}^{n/2}
\left ( \frac{a_1^2}{4} + (2k-1)^2 a_2 \right ) \right | 
\left [ 1 + O \left ( \frac{\log n }{ n }\right ) \right ],
\end{equation}
and for odd $n,$ 
\begin{equation}
\label{f19}
\gamma_n =
\left | \frac{8}{2^n [(n-1)!]^2}  \frac{a_1}{2}
\prod_{k=1}^{(n-1)/2}
\left ( \frac{a_1^2}{4} + (2k)^2 a_2 \right ) \right |
\left [ 1 + O \left ( \frac{\log n }{ n }\right ) \right ].
\end{equation}
\end{Theorem}

Observe, that Theorem \ref{thm4b} holds for $a_2 = 0 $ as well,
because then (\ref{f18}) and (\ref{f19}) come from
(\ref{i09}).
Of course, one can give an alternative direct proof of (\ref{f18})
and (\ref{f19}) in the case where $ a_2 = 0 $ by the same argument that
has been used to prove Theorem~\ref{thm4}.

If $K > 2, $ then, besides the simple extensions of
Theorem \ref{thm4b} that come from Proposition \ref{prop2},
we don't know any explicit
formula (in terms of the coefficients $a_k $ in (\ref{f1}))
for the exact asymptotics of $\gamma_n $ as $n \to \infty $
although some general formulas for the asymptotics in the case
of trig-polynomial potentials could be found, for example, in \cite{Gr}.

\end{document}